\documentclass[prb,twocolumn,aps,showpacs]{revtex4}
\usepackage{bm}
\usepackage{amsmath}
\usepackage{amssymb}
\usepackage[dvips]{color}
\usepackage[dvips]{graphics}
\usepackage[dvips]{graphicx}
\usepackage{eepic}
\usepackage{pstricks}%
\usepackage{pst-eps}%
\usepackage{pst-plot}%
\usepackage{pst-node}%
\renewcommand{\d}{{\mathrm d}}
\renewcommand{\i}{{\mathrm i}}
\newcommand{\e}{{\rm e}}

\newcommand{\tr}{{\rm tr}\;}
\renewcommand{\Re}{{\rm Re}\;}
\renewcommand{\Im}{{\rm Im}\;}
\newcommand{\sgn}{{\rm sgn}}
\renewcommand{\arctan}{{\rm Tan}^{-1}}
\newcommand{\arctanh}{{\rm Tanh}^{-1}}

\begin{document}
\title{Electric transport and magnetic properties in multilayer graphene}
 \author{Masaaki Nakamura$^1$ and  Lila Hirasawa$^{2,3}$}
\affiliation{
 $^1$Department of Applied Physics, Faculty of Science,
 Tokyo University of Science, 1-3 Kagurazaka, Shinjuku-ku,
 Tokyo 162-8601 Japan,\\
 $^2$Institute for Solid State Physics, University of Tokyo,
 Kashiwanoha, Kashiwa-shi, Chiba 277-8581 Japan,\\
 $^3$Department of Physics, Tokyo Institute of Technology,
 Oh-Okayama, Meguro-ku, Tokyo 152-8551 Japan}
\date{\today}

\begin{abstract}
 We discuss electric transport and orbital magnetism of multilayer
 graphenes in a weak-magnetic field using the matrix decomposition
 technique. At zero temperature, the minimum conductivity is given by
 that of the monolayer system multiplied by the layer number $N$,
 independent of the interlayer hopping $t$. When the interlayer hopping
 satisfies the condition $t\gg \hbar/\tau$ with $\tau$ being collision
 time of impurity scattering, $[N/2]$ kinks and $[N/2]+1$ plateaux
 appear in the Fermi-energy (gate voltage) dependence of the
 conductivity and the Hall conductivity, respectively.  These behaviors
 are interpreted as multiband effects. We also found that the Hall
 conductivity and the magnetic susceptibility take minimum value as a
 function of temperature, for certain value of the gate voltage. This
 behavior is explained by Fermi-energy dependence of these functions at
 zero temperature.
\end{abstract}

\maketitle

\section{Introduction}

Experimental studies of graphene have revealed exotic transport
properties such, as an anomalous quantum Hall effect and the finite
universal conductivity at zero energy.\cite{Novoselov,Zhang} These
results are essentially explained by the two-dimensional (2D) massless
Dirac equation which describes the low-energy band structure of graphene
around the gapless point.\cite{Ludwig-F-S-G,Shon-A,Zheng-A,
Gusynin-S_2005b,Gusynin-S_2006,Peres-G-C,Ziegler_2007}
Moreover, multilayer graphenes which consist of stacked few-layer
systems also attract attention.  For bilayer systems, quantum Hall
effect\cite{Novoselov,Zhang,McCann-F} and longitudinal
conductivity\cite{Nilsson-C-G-P,Koshino-A_2006,Cserti,Cserti-C-D} have
been studied.  For systems with more than three layers, electronic
structures are investigated experimentally\cite{Ohta,Bostwick,Zhou} and
theoretically.\cite{Latil-H,Partoens-P,Guinea-C-P} One of the most
interesting point of these multilayer systems would be variety of
stacking structures. A graphene is usually produced by micromechanical
cleavage of graphite, so that the stacking structure is considered to be
the Bernal type, since the natural graphite falls into this
category. However, production of graphene with other stacking types may
also be possible by recent epitaxial methods.  The difference of band
structure depending on the stacking types are
discussed,\cite{Latil-H,Guinea-C-P} and stability of the stacking
structures is also studied in terms of symmetry
arguments.\cite{Manes-G-V}

On the other hand, the magnetic susceptibility of carbon systems has
been studied for long times, before the discovery of graphene. It is
well known that three-dimensional graphite shows large diamagnetism, and
this has been explained theoretically by McClure. He showed that the
orbital diamagnetism appears with a delta function peak at the zero
energy point based on the 2D massless Dirac
equation.\cite{McClure,Safran-D} This argument can be essentially
applied to monolayer graphenes,\cite{Ghosal-G-C} and effects of impurity
scattering\cite{Fukuyama_2007,Koshino-A_2007a} and of an energy
gap\cite{Nakamura} are discussed. The orbital magnetism in multilayer
systems was studied more than two decades ago, motivated by graphite
intercalation compounds.  Especially, the magnetic susceptibility of
bilayer and multilayer systems was discussed by Safran,\cite{Safran} and
by Saito and Kamimura,\cite{Saito-K} respectively.  Quite recently,
Koshino and Ando calculated the susceptibility using matrix
decomposition technique.\cite{Koshino-A_2007b} They discussed that
Hamiltonian of the Bernal stacking systems can be block diagonalized
into effective bilayer and monolayer Hamiltonians depending on parity of
layer numbers. This is a powerful tool to investigate multilayer
systems.

In this paper, we turn our attention to the electric conductivity and
Hall conductivity of multilayer graphenes in a weak-magnetic field.  We
also consider how the differences of stacking structures appear in the
physical quantities. We use the matrix decomposition technique used by
Koshino and Ando throughout this paper for the calculation of Bernal
stacking systems. We also discuss the finite-temperature properties
including magnetic susceptibility in these systems.

The rest of paper is organized as follows. In Sec.~\ref{sec:Multilayer},
we discuss the Hamiltonian of the multilayer systems. In
Secs.~\ref{sec:Conductivity} and \ref{sec:Hall}, the conductivity and
the Hall conductivity are discussed, respectively. Kinks and plateaux
appearing in their Fermi-energy dependence are discussed.  In
Sec.~\ref{sec:Orbital_Magnetism}, we discuss the finite-temperature
properties of the diamagnetic orbital susceptibility, and discuss a
minimum value as a function of temperature. In Appendices, we discuss
the decomposition of Hamiltonian, and present analytical forms of
physical quantities.

\begin{figure*}
 \begin{center}
   \psset{unit=10mm}
   \begin{pspicture}(-1,-1)(2,2)
    \rput(2.0,-1.2){(a) Bernal stacking}
    \rput(-0.8,-0.2){$1$}
    \rput(-0.6,0.5){$\textcolor{blue}{2}$}
    \rput(-0.8,1.2){$1$}
    \def\tiltedhexagon{
    \pspolygon[linewidth=1.0pt]
    (-0.28867513,0.2)(0.28867513,0.2)(0.57735027,0)
    (0.28867513,-0.2)(-0.28867513,-0.2)(-0.57735027,0)}
    \rput(0,0){\tiltedhexagon}
    \rput(0,-0.4){\tiltedhexagon}
    \rput(0.86602540,-0.2){\tiltedhexagon}
    \rput(0.28867513,0.5){\tiltedhexagon}
    \pscircle*[linecolor=red](0.57735027,0){0.05}
    \pscircle*[linecolor=blue](0.28867513,-0.2){0.05}
    \pscircle*[linecolor=red](0,0.7){0.05}
    \pscircle*[linecolor=red](0,0.3){0.05}
    \pscircle*[linecolor=red](0.86602540,0.5){0.05}
    \pscircle*[linecolor=blue](0.57735027,0.7){0.05}
    \psline[linecolor=cyan]{-}(0.57735027,-0.4)(0.57735027,1.0)
    \psline[linecolor=cyan]{-}(-0.28867513,-0.2)(-0.28867513,1.2)
    \rput(0,1.4){
    \rput(0,0){\tiltedhexagon}
    \rput(0,-0.4){\tiltedhexagon}
    \rput(0.86602540,-0.2){\tiltedhexagon}
    \pscircle*[linecolor=red](0.57735027,0){0.05}
    \pscircle*[linecolor=red](0.57735027,-0.4){0.05}    
    \pscircle*[linecolor=red](-0.28867513,-0.2){0.05}        
    \pscircle*[linecolor=blue](0.28867513,-0.2){0.05}}
    \pscircle*[linecolor=blue](0.57735027,0.3){0.05}
    \pscircle*[linecolor=blue](-0.28867513,0.5){0.05}
    \pscircle*[linecolor=red](0.57735027,-0.4){0.05}
    \pscircle*[linecolor=red](-0.28867513,-0.2){0.05}
    \psline[linestyle=dotted](0.28867513,0.5)(0.28867513,1.2)
    \psline[linestyle=dotted](0,-0.37)(0,0.3)
    \psline[linestyle=dotted](0.86602540,-0.2)(0.86602540,0.5)   
\end{pspicture}
    \psset{unit=10mm}
   \begin{pspicture}(-1,-1)(4,3)
    \def\hexagon{
    \pspolygon[linewidth=1.0pt]
    (-0.28867513,0.5)(0.28867513,0.5)(0.57735027,0)
    (0.28867513,-0.5)(-0.28867513,-0.5)(-0.57735027,0)}
    \def\hexagonb{
    \pspolygon[linewidth=1.0pt,linecolor=blue,linestyle=dashed]
    (-0.28867513,0.5)(0.28867513,0.5)(0.57735027,0)
    (0.28867513,-0.5)(-0.28867513,-0.5)(-0.57735027,0)}
    \def\pairhexa{
    \rput(0,0){\hexagon}
    \rput(0.57735027,0){\hexagonb}}
    \rput(0,0){\pairhexa}
    \rput(0,1){\pairhexa}
    \rput(0.86602540,-0.5){\pairhexa}
    \rput(0.86602540,0.5){\pairhexa}
    \rput(0.86602540,1.5){\pairhexa}
    \rput(1.73205080,0){\pairhexa}
    \rput(1.74205080,1){\pairhexa}
   \end{pspicture}
%
   \psset{unit=10mm}
   \begin{pspicture}(-1,-1)(3,3)
    \rput(2.0,-1.2){(b) rhombohedral stacking}
    \rput(-0.8,-0.2){$1$}
    \rput(-0.51132487,0.7){$\textcolor{blue}{2}$}
    \rput(-0.22294973,1.6){$\textcolor{red}{3}$}
    \def\tiltedhexagon{
    \pspolygon[linewidth=1.0pt]
    (-0.28867513,0.2)(0.28867513,0.2)(0.57735027,0)
    (0.28867513,-0.2)(-0.28867513,-0.2)(-0.57735027,0)}
    \multirput(0,0)(0.28867513,0.9){3}{
    \rput(0,0){\tiltedhexagon}
    \rput(0,-0.4){\tiltedhexagon}
    \rput(0.86602540,-0.2){\tiltedhexagon}
    \pscircle*[linecolor=red](0.57735027,0){0.05}
    \pscircle*[linecolor=blue](0.28867513,-0.2){0.05}
    }
    \psline[linecolor=cyan]{-}(0.57735027,-0.4)(0.57735027,0.3)
    \psline[linecolor=cyan]{-}(-0.28867513,-0.2)(-0.28867513,0.5)
    \psline[linecolor=cyan]{-}(0.86602540,0.5)(0.86602540,1.2)
    \psline[linecolor=cyan]{-}(0,0.7)(0,1.4)
    \multirput(0,0)(0.28867513,0.9){3}{
    \pscircle*[linecolor=red](-0.28867513,-0.6){0.05}
    \pscircle*[linecolor=blue](0.28867513,-0.6){0.05}
    \pscircle*[linecolor=blue](-0.57735027,-0.4){0.05}
    \pscircle*[linecolor=red](0.57735027,-0.4){0.05}
    \pscircle*[linecolor=red](-0.28867513,-0.2){0.05}
    }
    \psline[linestyle=dotted](0.28867513,0.5)(0.28867513,1.2)
    \psline[linestyle=dotted](0,-0.4)(0,0.3)
    \psline[linestyle=dotted](0.86602540,-0.2)(0.86602540,0.5)   
    \psline[linestyle=dotted](1.15470053,0.7)(1.15470053,1.4)   
   \end{pspicture}
    \psset{unit=10mm}
   \begin{pspicture}(-1,-1)(4,3)
    \def\hexagon{
    \pspolygon[linewidth=1.0pt]
    (-0.28867513,0.5)(0.28867513,0.5)(0.57735027,0)
    (0.28867513,-0.5)(-0.28867513,-0.5)(-0.57735027,0)}
    \def\hexagonb{
    \pspolygon[linewidth=1.0pt,linecolor=blue,linestyle=dashed]
    (-0.28867513,0.5)(0.28867513,0.5)(0.57735027,0)
    (0.28867513,-0.5)(-0.28867513,-0.5)(-0.57735027,0)}
    \def\hexagonr{
    \pspolygon[linewidth=1.0pt,linecolor=red,linestyle=dashed,dash=1pt 1pt]
    (-0.28867513,0.5)(0.28867513,0.5)(0.57735027,0)
    (0.28867513,-0.5)(-0.28867513,-0.5)(-0.57735027,0)}
    \def\hexas{
    \rput(0,0){\hexagon}
    \rput(0.57735027,0){\hexagonb}
    \rput(0.28867513,0.5){\hexagonr}}
    \rput(0,0){\hexas} 
    \rput(0,1){\hexas} 
    \rput(1.73205080,0){\hexas}
    \rput(1.74205080,1){\hexas} 
    \psline[linewidth=1.0pt](0.28867513,1.5)
    (0.57735027,2)(1.1547005,2)(1.4433757,1.5)
    \psline[linewidth=1.0pt](0.28867513,-0.5)
    (0.57735027,-1)(1.1547005,-1)(1.4433757,-0.5)
    \psline[linewidth=1.0pt](0.57735027,0)(1.1547005,0)
    \psline[linewidth=1.0pt](0.57735027,1)(1.1547005,1) 
    \def\bluelines{
    \psline[linewidth=1.0pt,linecolor=blue,linestyle=dashed]
    (0.28867513,1.5)(0.57735027,2)(1.1547005,2)(1.4433757,1.5)
    \psline[linewidth=1.0pt,linecolor=blue,linestyle=dashed]
    (0.28867513,-0.5)(0.57735027,-1)(1.1547005,-1)(1.4433757,-0.5)
    \psline[linewidth=1.0pt,linecolor=blue,linestyle=dashed]
    (0.57735027,0)(1.1547005,0)
    \psline[linewidth=1.0pt,linecolor=blue,linestyle=dashed]
    (0.57735027,1)(1.1547005,1) 
    } 
    \def\redlines{
    \psline[linewidth=1.0pt,linecolor=red,linestyle=dashed,dash=1pt 1pt]
    (0.28867513,1.5)(0.57735027,2)(1.1547005,2)(1.4433757,1.5)
    \psline[linewidth=1.0pt,linecolor=red,linestyle=dashed,dash=1pt 1pt]
    (0.28867513,-0.5)(0.57735027,-1)(1.1547005,-1)(1.4433757,-0.5)
    \psline[linewidth=1.0pt,linecolor=red,linestyle=dashed,dash=1pt 1pt]
    (0.57735027,0)(1.1547005,0)
    \psline[linewidth=1.0pt,linecolor=red,linestyle=dashed,dash=1pt 1pt]
    (0.57735027,1)(1.1547005,1) 
    }
    \rput(0.57735027,0){\bluelines}
    \rput(0.28867513,0.5){\redlines}
   \end{pspicture}
\end{center}
 \caption{(Color online) Two stacking structures of multilayer graphene.}
 \label{fig:stackings}
 \begin{center}
  \includegraphics[height=55mm]{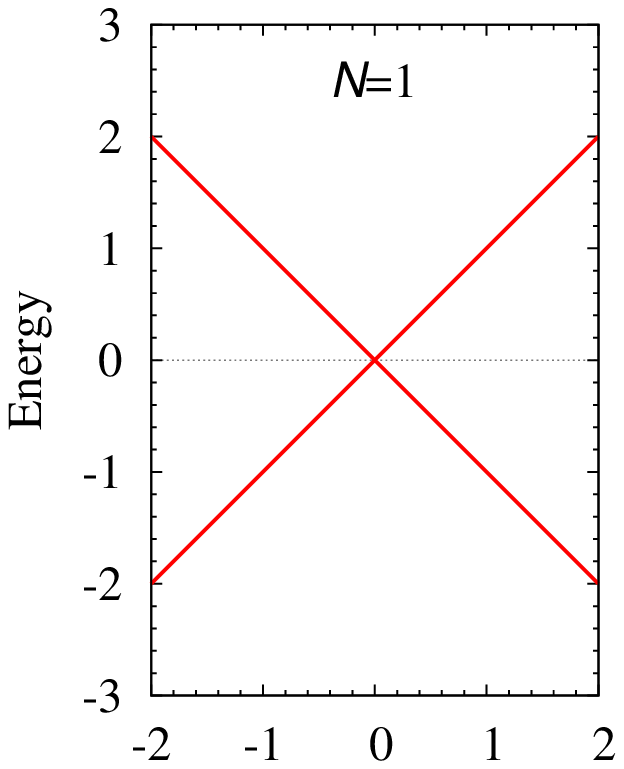}%
  \includegraphics[height=55mm]{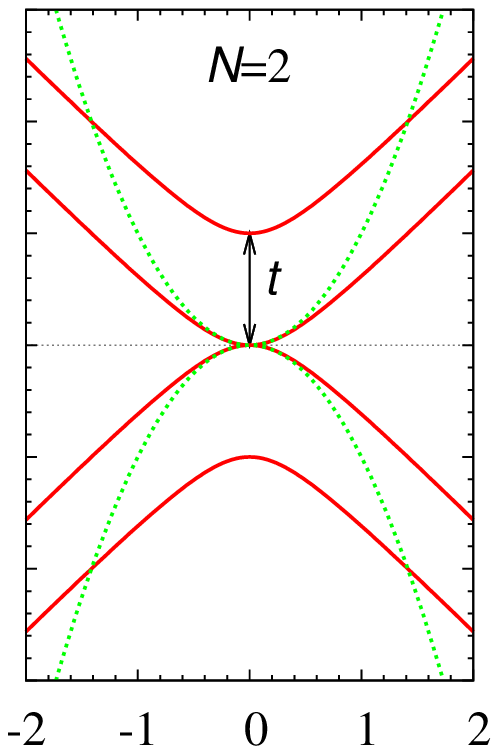}%
  \includegraphics[height=55mm]{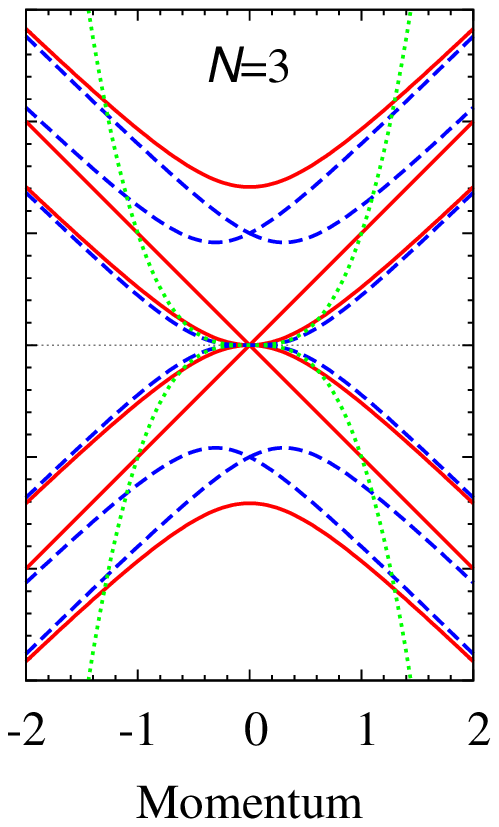}%
  \includegraphics[height=55mm]{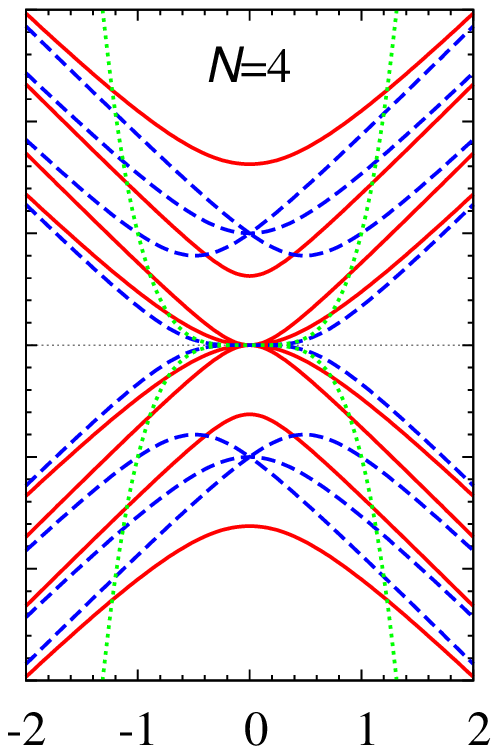}%
  \includegraphics[height=55mm]{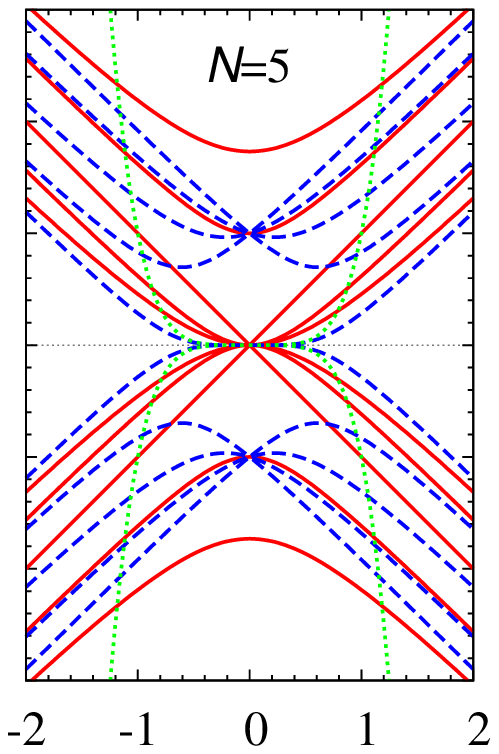}
 \caption{(Color online) Band structure of multilayer systems
  ($N=1$-$5$) with the Bernal stacking [Eq.~(\ref{Ham_Bernal}), solid
  line] and with the rhombohedral stacking
  [Eq.~(\ref{Ham_rhombohedral}), dashed line]. For the Bernal stacking,
  linear dispersion relations appear for odd layer cases.  The
  dispersion relation of the effective Hamiltonian of the rhombohedral
  stacking [Eq.~(\ref{eff_Ham_rhombohedral}), dotted line] is also
  shown.}  \label{fig:bands}
 \end{center}
\end{figure*}

\section{Stacking structures
}\label{sec:Multilayer}

As is well known, the band structure of graphene has two gapless points
due to the hexagonal lattice, and the low-energy property of this system
is described by the 2D massless Dirac equation. For the multilayer
graphene, we consider network of the Dirac fermion systems connected by
interlayer hopping $t$. It is also known that there are mainly two
stacking types for graphite: one is the Bernal (staggered) stacking
where the layer sequence can be written $1212\dots$, and the other is
rhombohedral stacking $123123\dots$.  (see Fig.~\ref{fig:stackings})

The Hamiltonians of the Bernal stacking systems are given by
\begin{equation}
 \mathcal{H}_{\rm b}
=\left[
  \begin{array}{ccc ccc ccc c}
   & \tilde{\pi}_- & & t & & & & & &\\
   \tilde{\pi}_+ & & & & & & & & &\\
   & & & \tilde{\pi}_-  & & & & & &\\
   t & & \tilde{\pi}_+ & & t & & & & &\\
   & & & t & & \tilde{\pi}_-  & & t & &\\
   & & & & \tilde{\pi}_+  & & & & &\\
   & & & & & & & \tilde{\pi}_-  & &\\
   & & & & t & & \tilde{\pi}_+ & & t &\\
   & & & & & & & t & &\\
   \\
  \end{array}
\right],
\label{Ham_Bernal}
\end{equation}
where $\tilde{\pi}_{\pm}\equiv v\pi_{\pm}$ with $v$ being the velocity
of the Dirac equation for the monolayer system, and
$\pi_{\pm}\equiv\pi_x\pm\i\pi_y$.  $\bm{\pi}\equiv\bm{p}+e\bm{A}/c$ is
the momentum operator in a magnetic field $\nabla\times\bm{A}=(0,0,B)$.
The band structures obtained as eigenvalues of this Hamiltonian are
shown in Fig.~\ref{fig:bands}.  Quite recently, Koshino and Ando proved
that the effective Hamiltonian of $N$-layer Bernal stacking system
becomes, without loss of generality, isolated $[N/2]$ bilayer system
($[x]$ is the maximum integer which does not exceed $x$) with effective
interlayer hoppings $t^*_1,t^*_2,\dots,t^*_{[N/2]}$, and one monolayer
system if $N$ is odd.\cite{Koshino-A_2007b} This result can also be
obtained by calculating determinant of the Schr\"{o}dinger equation.
(see Appendix~\ref{sec:eff_Ham}) The values of effective hoppings are
given in Table~\ref{EILH}.  We will use this decomposed effective
Hamiltonian throughout this paper. For the bilayer system, an energy gap
$\Delta=t$ between two bands appears, since the energy spectra are given
by $\varepsilon=\pm(\sqrt{t^2+(2v\hbar k)^2}\pm t)/2$.

On the other hand, the rhombohedral stacking is described by
\begin{equation}
 \mathcal{H}_{\rm r}
=\left[
  \begin{array}{ccc ccc ccc c}
   & \tilde{\pi}_- & & t & & & & & &\\
   \tilde{\pi}_+ & & & & & & & & &\\
   & & & \tilde{\pi}_-  & & t & & & &\\
   t & & \tilde{\pi}_+ & & & & & & &\\
   & & &  & & \tilde{\pi}_-  & & t & &\\
   & & t & & \tilde{\pi}_+  & & & & &\\
   & & & & & & & \tilde{\pi}_-  & &\\
   & & & & t & & \tilde{\pi}_+ & &  &\\
   & & & & & & & & &\\
   \\
  \end{array}
\right].
\label{Ham_rhombohedral}
\end{equation}
The band structure of this system is complicated, as shown in
Fig.~\ref{fig:bands}, and decomposition such as the Bernal stacking is
difficult. If we turn our attention to the two bands near the
zero-energy point, the effective Hamiltonian becomes the following
$2\times 2$ form\cite{McCann-F,Manes-G-V} (see
Appendix~\ref{sec:eff_Ham})
\begin{equation}
 {\cal H}_{\rm eff}=-\frac{1}{t^{N-1}}\left[
  \begin{array}{cc}
    0   &(v\pi_+)^N    \\
    (v\pi_-)^N   & 0
  \end{array}
\right].
\label{eff_Ham_rhombohedral}
\end{equation}
This effective Hamiltonian is useful to discuss the quantum Hall
effect\cite{McCann-F} and the zero-energy longitudinal
conductivity,\cite{Cserti,Cserti-C-D} but it is difficult to obtain
physically relevant results for the Hall conductivity and the
susceptibility in the weak-magnetic field treatment below (see
Appendix~\ref{sec:Analytic_results}). Therefore, calculations in this
paper is mainly devoted to Bernal stacking systems.

The low-energy effective Hamiltonian (\ref{eff_Ham_rhombohedral}),
however, gives information of the reason why the Bernal stacking is more
stable than the rhombohedral stacking: Since the dispersion relation is
$\varepsilon=\pm(v\hbar k)^N/t^{N-1}$, the density of states of the
$N$-layer rhombohedral system is given by
\begin{equation}
 D_N(\varepsilon)=\frac{V t}{2\pi N \hbar^2 v^2}
  \left(\frac{|\varepsilon|}{t}\right)^{2/N-1},
\end{equation}
where $V$ is volume of the system.  For $N\geq 3$, $D_N(\varepsilon)$
diverges at $\varepsilon=0$, where the density of states of the Bernal
case is $D_2(\varepsilon)=$ const.  Therefore, the rhombohedral systems
are considered to be unstable against external perturbations.

\begin{table}
\begin{ruledtabular}
\begin{tabular}{|c|c|c|c|c|}
 $N$ & $t_1^*/t$ & $t_2^*/t$ & $t_3^*/t$ & $\cdots$\\
 \hline
 $2$ & $1$       &   ---   & --- &\\
 $3$ & $\sqrt{2}$ &   ---   & --- &\\
 $4$ & $\frac{\sqrt{5}-1}{2}$ & $\frac{\sqrt{5}+1}{2}$ & --- &\\
 $5$ & $1$       & $\sqrt{3}$ & --- &
\end{tabular}
\caption{Effective interlayer hopping integral $t_i^*$
 ($i=1,2,\dots,[N/2]$) of the effective bilayer Hamiltonian of the
 Bernal stacking systems.}\label{EILH}
\end{ruledtabular}
\end{table}

\section{Conductivity}\label{sec:Conductivity}

First, we consider the conductivity of the multilayer systems based on
the linear response theory.  The conductivity is given by the Kubo
formula\cite{Mahan}
\begin{equation}
 \Re\sigma_{\mu\nu}
  =\lim_{\omega\to 0}
  \frac{\Im\tilde{\Pi}_{\mu\nu}(\bm{0},\omega+\i 0)}{\hbar\omega},
   \label{Kubo_form}
\end{equation}
where $\tilde{\Pi}_{\mu\nu}(\bm{q},\omega)\equiv
\Pi_{\mu\nu}(\bm{q},\omega)-\Pi_{\mu\nu}(\bm{q},0)$.  The polarization
function in the Matsubara form is given by
\begin{align}
 &
 \Pi_{\mu\nu}(\bm{q},\i\nu_m)
   =\frac{1}{V}\int_{0}^{\beta\hbar}\d\tau\e^{\i\nu_m\tau}
 \langle{\cal T}_{\tau}
 J_{\mu}(\bm{q},\tau)J_{\nu}(\bm{0},0)\rangle\label{Pi_CC},\\
 &
 \tilde{\Pi}_{\mu\nu}(\bm{0},\i\nu_m)=
 -\frac{e^2}{\beta\hbar V}
 \sum_{\bm{k}}\sum_{n}
 \tr\left({\cal G}\gamma_{\mu}{\cal G}_+\gamma_{\nu}\right),
 \label{Pi_CC2}
\end{align}
where $\beta\equiv 1/k_{\rm B}T$ and $V$ are the inverse temperature and
the volume of the system, respectively. The current operator is given by
\begin{equation}
 J_{\mu}(-\bm{q},\tau)=
  -e\sum_{\bm{k}}
  \Psi^{\dag}(\bm{k}+\bm{q}/2,\tau)
  \gamma_{\mu}\Psi(\bm{k}-\bm{q}/2,\tau),
\end{equation}
where $\Psi(\bm{k},\tau)$ is the Fourier component of the field operator
$\Psi^{\dag}(\bm{r})=[ \psi^{\dag}_{A_1}(\bm{r}),
\psi^{\dag}_{B_1}(\bm{r}), \psi^{\dag}_{A_2}(\bm{r}),
\psi^{\dag}_{B_2}(\bm{r}),\dots ]$, where $A_i$ and $B_i$ indicate two
sublattices of the hexagonal lattice of $i$th layer. The matrix
$\gamma_{\mu}$ is defined by
\begin{equation}
 \gamma_{\mu} \equiv \frac{1}{\hbar}\frac{\partial{\cal H}}{\partial
  k_{\mu}}.
  \label{def_gamma}
\end{equation}
In Eq.~(\ref{Pi_CC2}), the impurity-averaged temperature Green function
is given by
\begin{equation}
 {\cal G}
  \equiv {\cal G}(\bm{k},\i\omega_n)
  =(\i\omega_n + [\mu + \i\ \sgn(\omega_n)\Gamma 
  -{\cal H}_0]/\hbar)^{-1},
\end{equation}
where $\omega_n\equiv(2n+1)\pi/\beta\hbar$ is the Matsubara frequency of
fermions. For ${\cal G}_+\equiv{\cal G}(\bm{k},\omega_n+\nu_m)$,
$\nu_m=2\pi m/\beta\hbar$ is the Matsubara frequency of bosons. Here, we
have introduced the scattering rate phenomenologically as the
quasiparticle self-energy $\Gamma=-\Im\Sigma^{\rm R}$, neglecting the
frequency and the momentum dependences.  This parameter stems from the
scattering of the impurity potential implicitly assumed in the present
system, and is related to the mean free time of quasiparticles as
$\Gamma=\hbar/2\tau$.\cite{Abrikosov-G-D}

The final analytic form of $\sigma_{xx}$ for a bilayer system is
presented in Appendix~\ref{sec:Analytic_results}.  In Fig.~\ref{fig.1},
the zero-temperature conductivity (calculated per valley and per spin)
$\sigma_{xx}$ versus the Fermi energy $\mu$ scaled by the scattering
rate $\Gamma$ is shown. Experimentally, $\mu$ is a tunable parameter by
the gate voltage. At zero energy $\mu/\Gamma=0$, the conductivity takes
the minimum value $\sigma_{xx}=Ne^2/\pi h$ which is the $N$ times of the
minimum conductivity of the monolayer
system.\cite{Ludwig-F-S-G,Shon-A,Gusynin-S_2005b,Gusynin-S_2006} For
this finite value $\sigma_{\rm min}=e^2/\pi h$, there are controversial
arguments,\cite{Ziegler_2007} but all the analytic results do not
coincide with the experimental value $e^2/h$,\cite{Novoselov,Zhang}
except for the numerical analysis of Nomura and
MacDonald.\cite{Nomura-M}

As the Fermi energy increases as $\mu/\Gamma\to\infty$ with small
interlayer hopping $t/\Gamma\ll 1$, the conductivity approaches to $N$
times of the Drude-Zener form,
\begin{equation}
 \sigma_{xx}=\frac{\sigma_0}{1+(\omega_{\rm c} \tau)^2}
  \simeq \sigma_0
  =\frac{e^2 \tau \mu}{4 \pi \hbar^2}.
 \label{sxx_DZ}
\end{equation} 
Compared with the result of the linearized Boltzmann equation for
nonrelativistic electrons $\sigma_0 = ne^2 \tau/m$, where $n$ is the
electron density, the ``electron mass'' $m$ is related as
$m=|\mu|/v^2$.\cite{Gusynin-S_2006} This agreement with the Boltzmann
description is due to the constant $\Gamma$. A similar linear
Fermi-energy dependence of the conductivity is reproduced by numerical
calculation considering effects of screened Coulomb impurity
scattering.\cite{Nomura-M}

For large interlayer hopping $t/\Gamma\gg 1$ which means that the energy
gap $\Delta \sim O(t)$ satisfies the condition $\Delta\gg \hbar/\tau$,
kinks appear in the Fermi-energy dependence. The number of kinks is
increased as the layers are increased. These phenomena can be
interpreted as a multiband effect: When the interlayer hopping is large
enough in the $N$-layer system, $N$ bands in the positive (negative)
energy region split into $[(N+1)/2]$ gapless modes and other bands with
different energy gaps. Therefore, the number of kinks in the positive or
negative energy region is $N/2$ for even layers, and $(N-1)/2$ for odd
layers, reflecting the discontinuity of the density of states.

In order to clarify this argument, we generalize Eq.~(\ref{sxx_DZ}) to
the multiband system for $\mu\geq 0$ as
\begin{equation}
 \sigma_{xx}=\frac{e^2\tau}{4 \pi\hbar}
  \sum_{i=1}^{N} \theta(\mu-v_i k_{{\rm F},i})\,
  v_i k_{{\rm F},i},
\end{equation}
where the velocity $v_i$ and the Fermi wave number of the $i$-th band
$k_{{\rm F},i}$ are related as
\begin{equation}
 mv_i=\hbar k_{{\rm F},i},\qquad \mu=\varepsilon_i(k_{{\rm F},i}).
  \label{e_mass}
\end{equation}
In Fig.~\ref{fig.2}, the Fermi-energy dependence of the zero-temperature
conductivity at $t/\Gamma=6$ for various layer numbers is shown. The
results of semiclassical analysis show good agreement with those of the
linear response theory, except for the small energy region where the
quantum effect is essential.  For more detailed semiclassical argument,
we should consider the Boltzmann equation for Dirac type
systems.\cite{Peres-LS-S}

\begin{figure}[h!]
 \begin{center}
  \includegraphics[width=75mm]{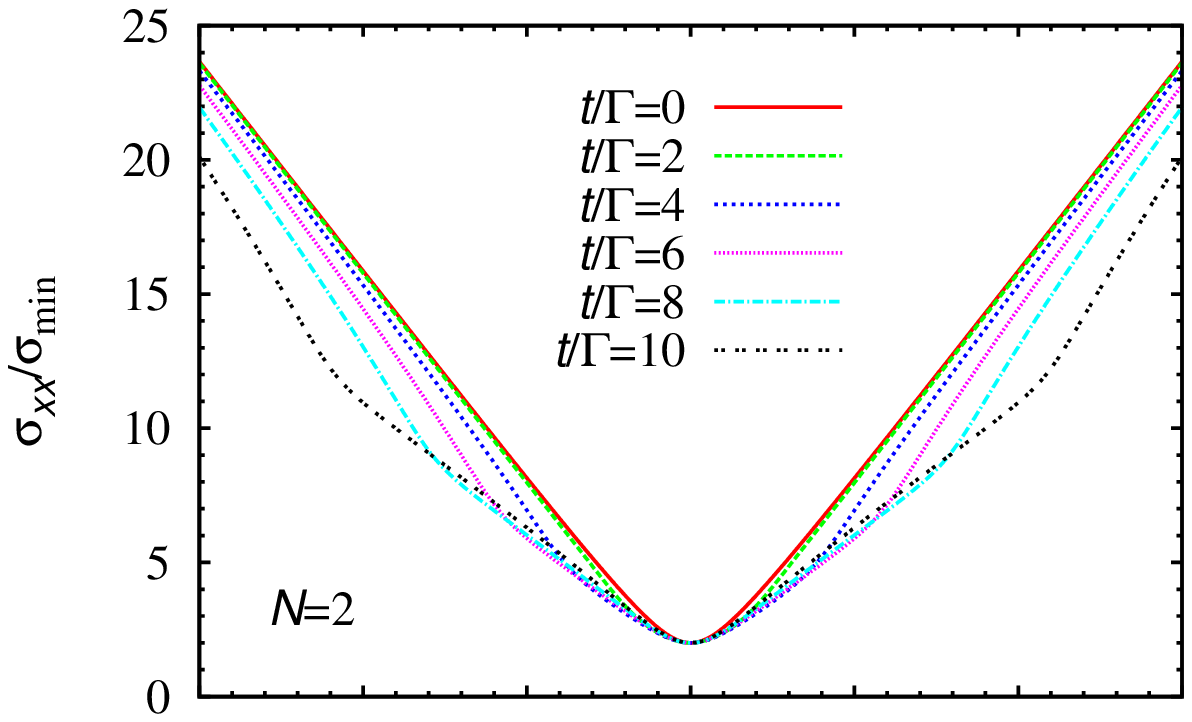}\\
  \includegraphics[width=75mm]{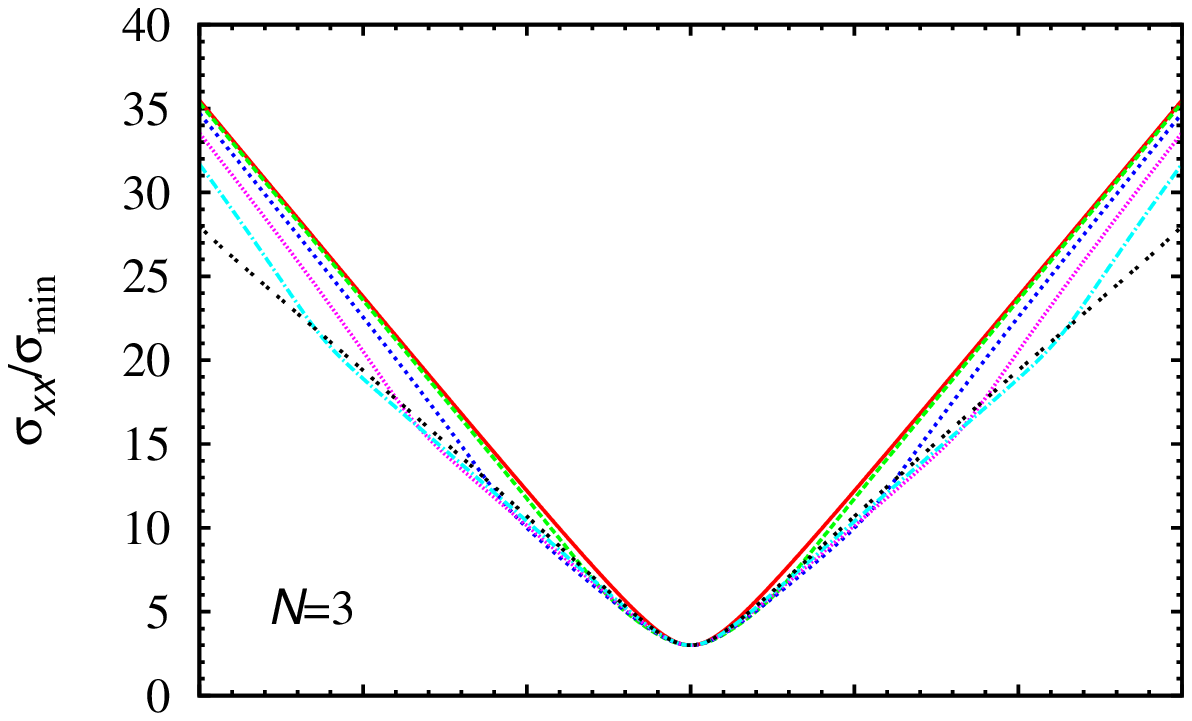}\\
  \includegraphics[width=75mm]{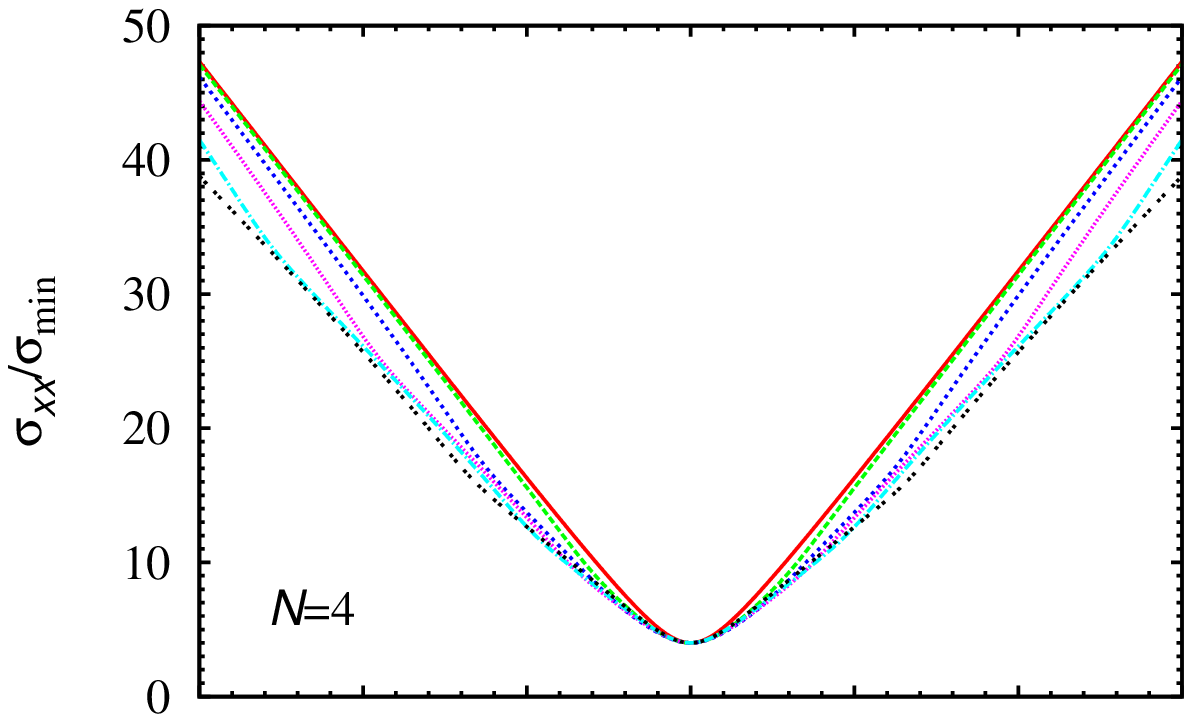}\\  
  \includegraphics[width=75mm]{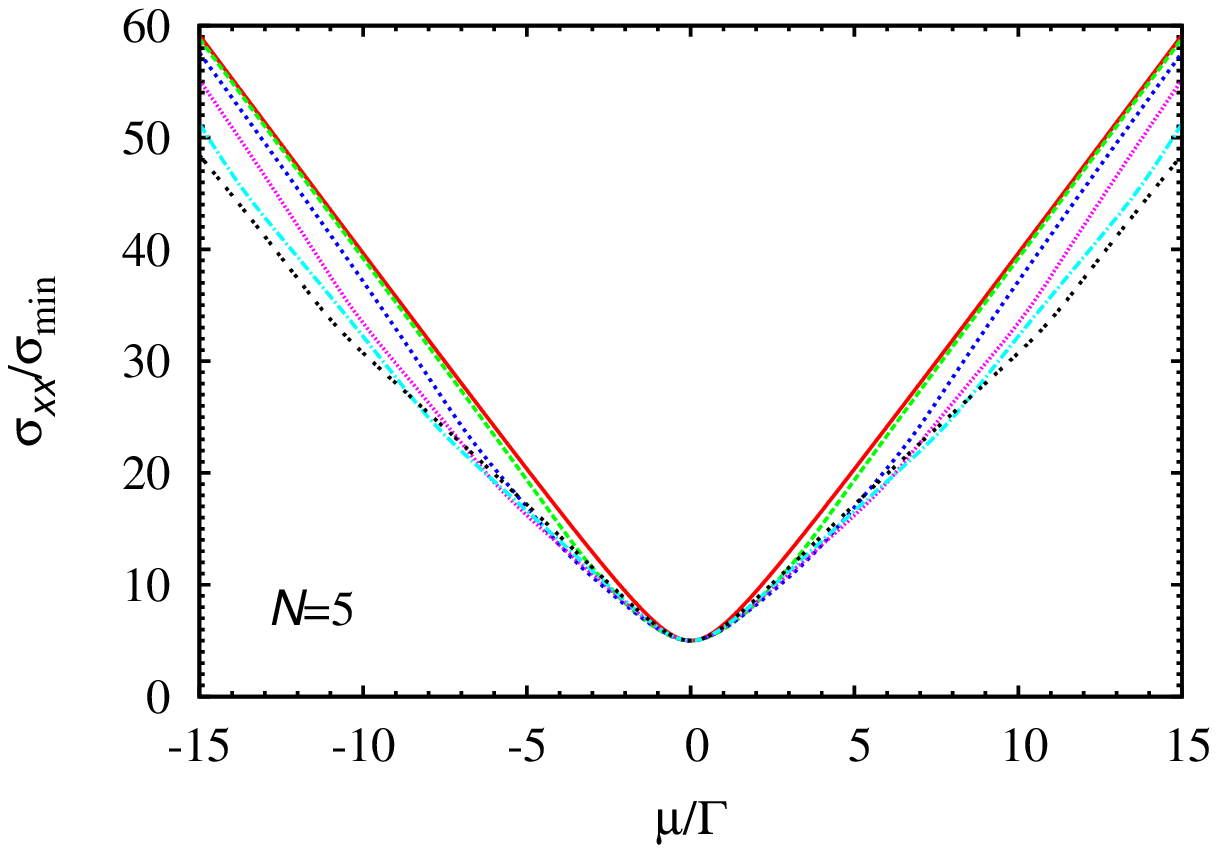}
 \end{center}
 \caption{(Color online) Fermi-energy ($\mu/\Gamma$) dependence of the
 longitudinal conductivity of Bernal stacking systems with $N=2,3,4,5$,
 scaled by the minimum conductivity of the monolayer system $\sigma_{\rm
 min}=e^2/\pi h$. In these results, $[N/2]$ kinks appear for the
 positive (or negative) Fermi-energy regions.}\label{fig.1}
\end{figure}

\begin{figure}
 \begin{center}
  \includegraphics[width=75mm]{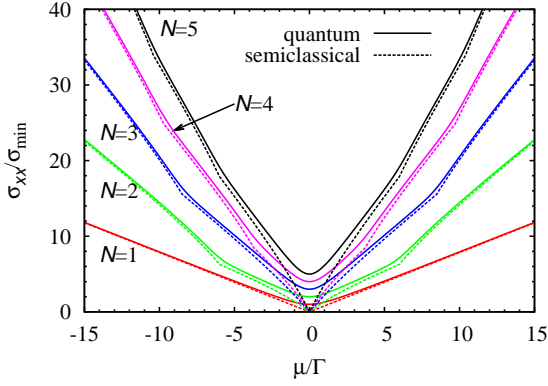}\\
 \end{center}
 \caption{(Color online) Comparison between the longitudinal
 conductivity obtained by the linear response theory (solid lines) and
 that by the semiclassical analysis (dashed lines) for one to five layer
 systems with $t/\Gamma=6$.}\label{fig.2}
\end{figure}

\section{Hall Effect}\label{sec:Hall}

Next, we consider the Hall conductivity $\sigma_{xy}$ in a weak-magnetic
field.  The expression of the Hall conductivity in terms of the Green
function is obtained by the Luttinger-Kohn
representation\cite{Luttinger-K} for the basic functions and the Fourier
expansion of the vector potential $\bm{A}(\bm{r})=\bm{A}_{\bm{q}}
\e^{\i\bm{q}\cdot\bm{r}}$. As the first-order perturbation of the
current term of the Hamiltonian ${\cal
H}-\bm{A}_{\bm{q}}\cdot\bm{J}(-\bm{q})/c$, we have the following three
point function:\cite{Fukuyama-E-W,Fukuyama_2006,Yang-N}
\begin{align}
 &
 \Pi_{\mu\nu}(\bm{q},\i\nu_m)=
 \sum_{\alpha=x,y}\frac{A_{\bm{q}\alpha}}{c\hbar}\frac{1}{V}
 \int_{0}^{\beta\hbar}\d\tau
 \int_{0}^{\beta\hbar}\d\tau'\nonumber\\
 &\times
 \e^{\i\nu_m\tau}
 \langle{\cal T}_{\tau}
 J_{\mu}(\bm{q},\tau)J_{\alpha}(-\bm{q},\tau')J_{\nu}(\bm{0},0)\rangle
 \label{Pi_CC3}.
\end{align}
Then by $\bm{q}$ expansion of the temperature Green function with the
relation $\partial_{k_{\mu}}{\cal G}={\cal G}\gamma_{\mu}{\cal G}$, we
obtain the polarization function in the linear order of the magnetic
field $B=\i(q_x A_{\bm{q}}^{y}-q_y A_{\bm{q}}^{x})$ as
\begin{align}
 &
 \tilde{\Pi}_{\mu\nu}(\bm{0},\i\nu_m)
 =\i\frac{e^3B}{2c\hbar}
 \frac{1}{V\beta\hbar}
 \sum_{\bm{k},n}\nonumber\\
 &\times
 \tr(
 \gamma_{\mu}{\cal G}_+\gamma_x{\cal G}_+\gamma_{\nu}{\cal G}\gamma_y{\cal G}
 -\gamma_{\mu}{\cal G}_+\gamma_y{\cal G}_+\gamma_{\nu}{\cal G}\gamma_x{\cal G}
 \nonumber\\
 &
+\gamma_{\mu} {\cal G}_+\gamma_x {\cal G}_+
\gamma_y {\cal G}_+\gamma_{\nu}{\cal G}
-\gamma_{\mu} {\cal G}_+\gamma_y {\cal G}_+
\gamma_x {\cal G}_+\gamma_{\nu} {\cal G}
 \nonumber\\
 &
+
 \gamma_{\mu}{\cal G}_+\gamma_{\nu}{\cal G}\gamma_x{\cal G}\gamma_y{\cal G}
-
 \gamma_{\mu}{\cal G}_+\gamma_{\nu}{\cal G}\gamma_y {\cal G}\gamma_x{\cal G}
 ).\label{Pi_CC4}
\end{align}

\begin{figure}[h!]
 \begin{center}
  \includegraphics[width=75mm]{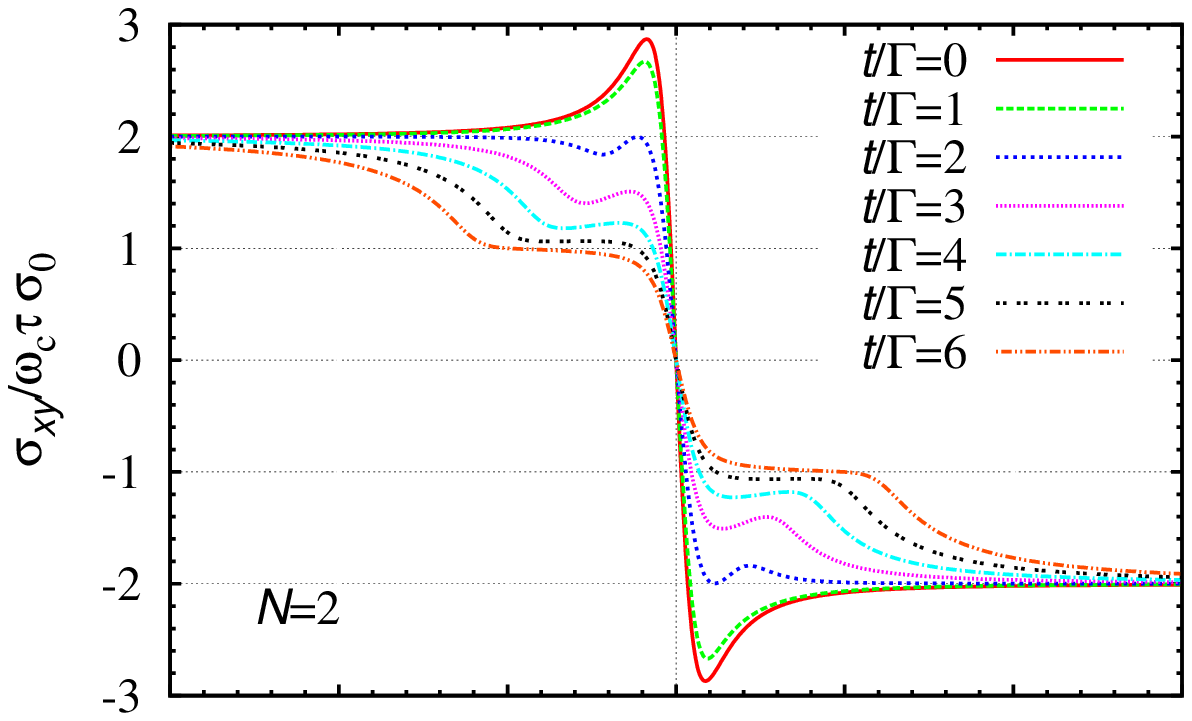}\\
  \includegraphics[width=75mm]{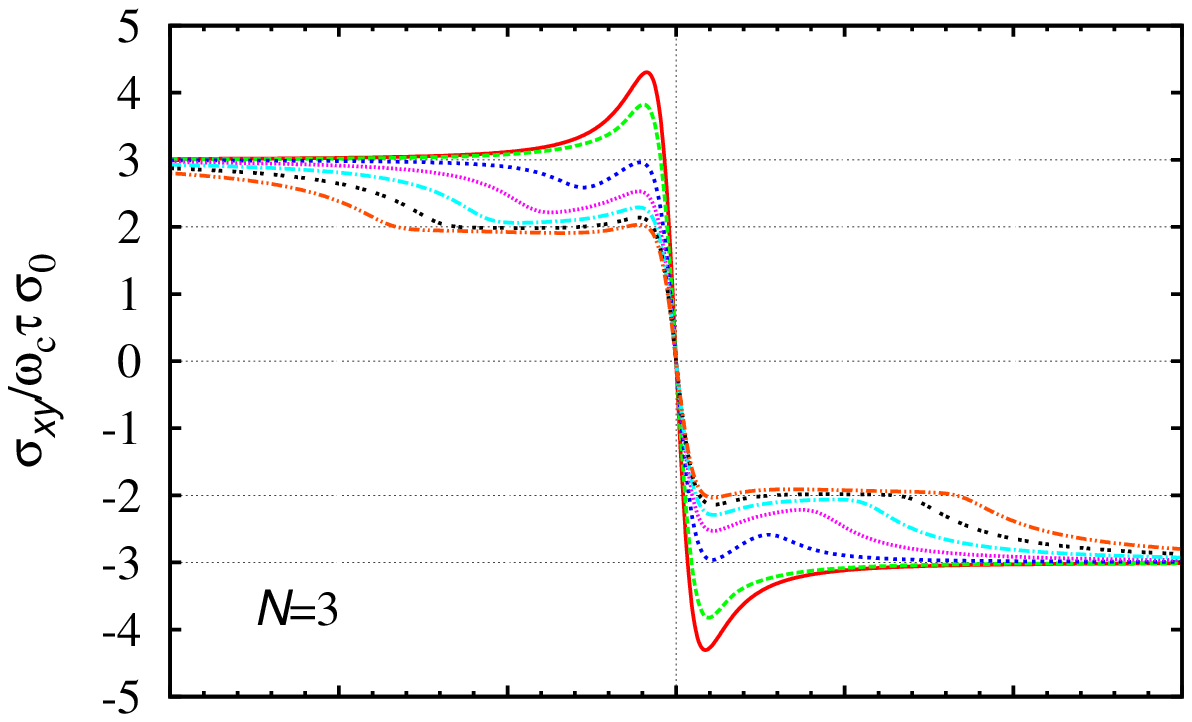}\\
  \includegraphics[width=75mm]{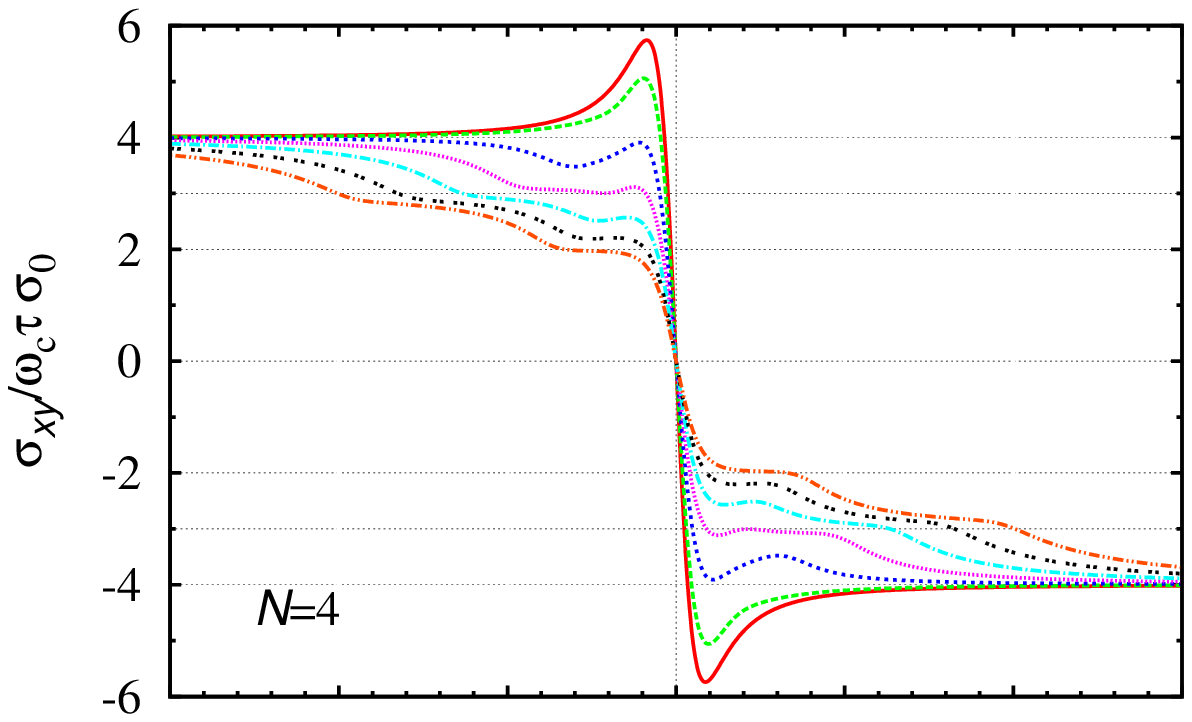}\\  
  \includegraphics[width=75mm]{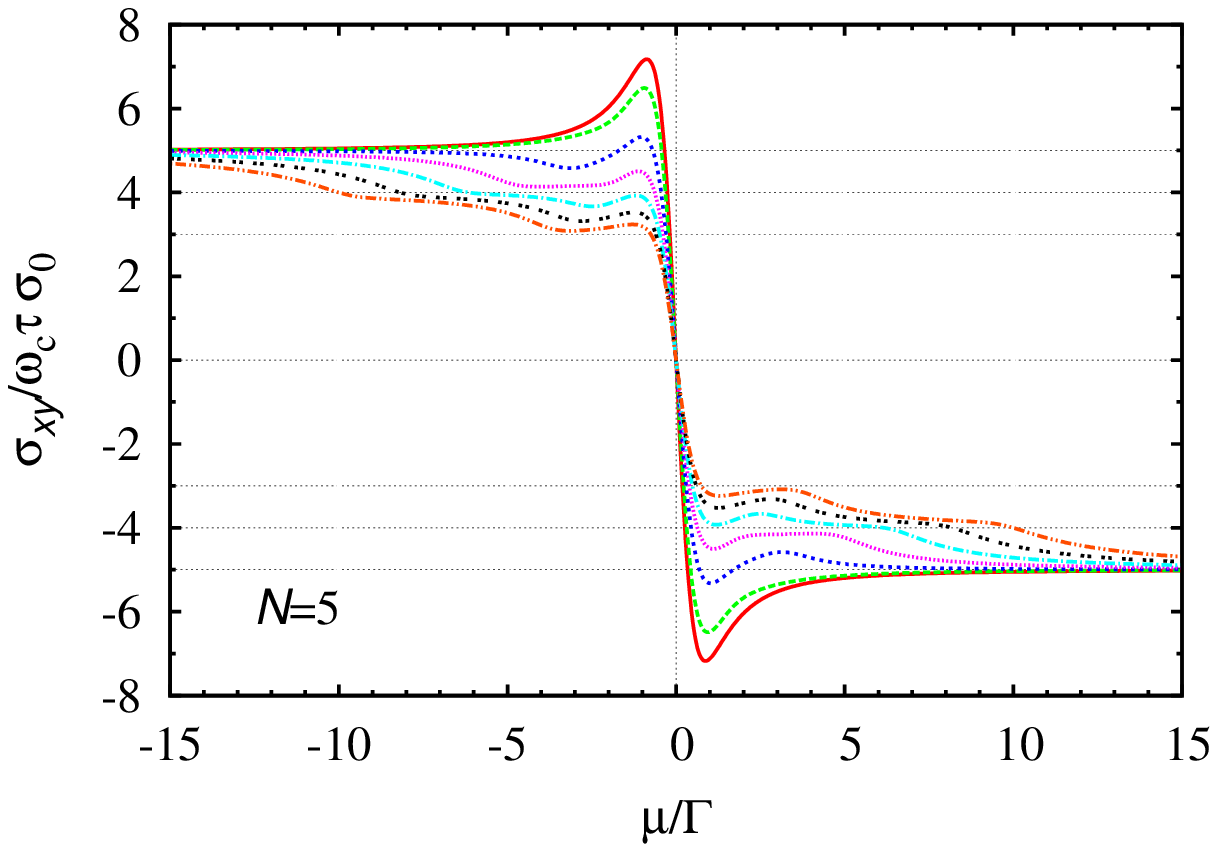}
 \end{center}
 \caption{(Color online) Fermi-energy ($\mu/\Gamma$) dependence of the
 Hall conductivity of Bernal stacking systems with $N=2,3,4,5$, scaled
 by the classical Hall conductivity of the monolayer system $\omega_{\rm
 c}\tau\sigma_0$. In these results, $[N/2]+1$ plateaux appear in large
 $t/\Gamma$ cases.}\label{fig.3}
\end{figure}

In Fig.~\ref{fig.3}, we show the zero-temperature Hall conductivity
(calculated per valley and per spin) $\sigma_{xy}$ versus the Fermi
energy (gate voltage) $\mu$ scaled by the scattering rate $\Gamma$.  The
Hall conductivity changes the sign depending on the sign of $\mu$.  For
small $t/\Gamma$, the Hall conductivity shows sharp change in the small
energy region $\mu/\Gamma\to 0$. Then its absolute value takes maximum,
and approaches to the constant value ($\propto\Gamma^{-2}$), as the
Fermi energy is increased as $\mu/\Gamma\rightarrow \infty$. This value
is $N$ times of the Drude-Zener like formula,
\begin{equation}
 \sigma_{xy}=
  \frac{-\omega_{\rm c}\tau\sigma_0}{1+(\omega_{\rm c}\tau)^2}
  \simeq -\omega_{\rm c}\tau\sigma_0,
\end{equation}
where $\omega_{\rm c}=|eB|/mc$ is the cyclotron frequency. The deviation
of the Hall conductivity from this classical value at the low-energy
regions and change of the sign are often called ``anomalous'' Hall
effect. Actually, this behavior is observed experimentally in the gate
voltage dependence of hall coefficient
$\rho_{xy}=-\sigma_{xy}/[(\sigma_{xx})^2+(\sigma_{xy})^2]$ at the
surface of a graphite.\cite{Morozov}

\begin{figure}
 \begin{center}
  \includegraphics[width=75mm]{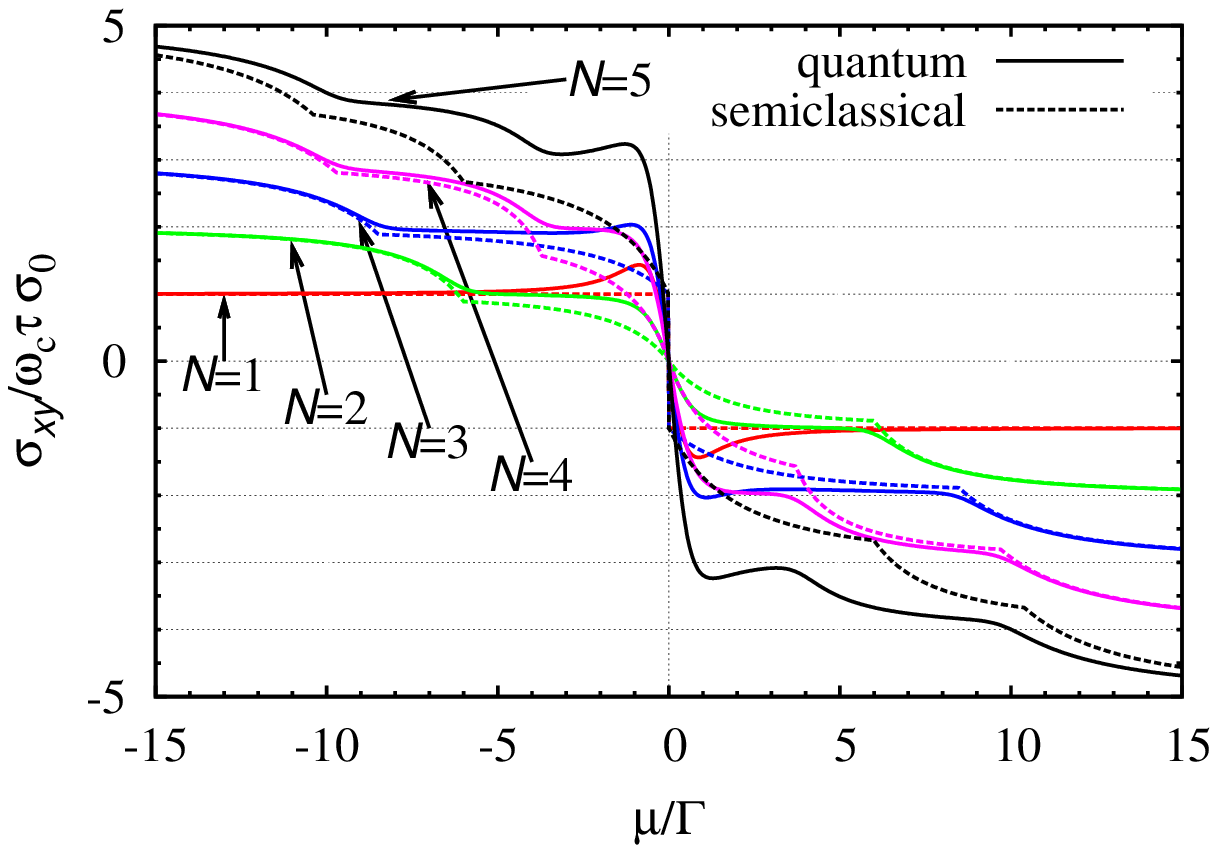}
 \end{center}
 \caption{(Color online) Comparison between the Hall conductivity
 obtained by the linear response theory (solid lines) and that by the
 semiclassical analysis (dashed lines) for one to five layer systems
 with $t/\Gamma=6$.}\label{fig.4}
\end{figure}

The most remarkable feature of the Hall conductivity is that plateaux
appear when the interlayer hopping is large $t/\Gamma\gg 1$.  The number
of plateaux for the positive or negative energy regions is $(N+2)/2$ for
even layers, and $(N+1)/2$ for odd layers, and $\sigma_{xy}$ is
``quantized'' by the unit of $\omega_{\rm c}\tau\sigma_0$.  This
phenomenon is due to effect of the energy gap $\Delta \sim O(t)$
satisfying the condition $\Delta\gg \hbar/\tau$.  Actually, a plateau
also appears in the result of the monolayer system with an energy
gap.\cite{Nakamura} In order to clarify this behavior, we also
generalize the Drude-Zener theory for $\sigma_{xy}$ in the multiband
systems. Then the Hall conductivity for $\mu\geq 0$ is given by
\begin{equation}
 \sigma_{xy}=-\omega_{\rm c}\tau\sigma_{xx}
  \simeq -\frac{e^3\tau^2 B}{4\pi c \hbar^2}
  \sum_{i=1}^{N} \theta(\mu-v_i k_{{\rm F},i})\, v_i^2,
\end{equation}
where the velocity $v_i$ and ``electron mass'' $m$ are given by
Eq.~(\ref{e_mass}). In Fig.~\ref{fig.4}, the Fermi-energy dependence of
the zero-temperature Hall conductivity $\sigma_{xy}$ at $t/\Gamma =6$
for various layer numbers is shown. The result of the semiclassical
analysis shows good agreement with those of the linear response theory
in terms of the step structure of $\sigma_{xy}$. For odd layers, the
peak structure remains even when the plateaux appear, because of the
contribution from a gapless monolayer mode.

\begin{figure}
 \begin{center}
  \includegraphics[width=75mm]{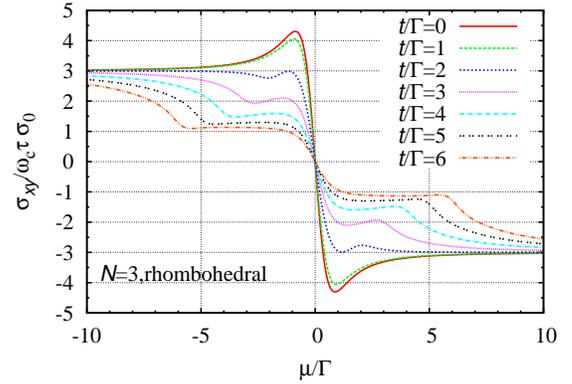}
 \end{center}
 \caption{(Color online) Fermi-energy ($\mu/\Gamma$) dependence of the
 Hall conductivity of a rhombohedral stacking system with $N=3$. In this
 case, the first plateau appears at $\sigma_{xy}\simeq\pm \omega_{\rm
 c}\tau\sigma_0$, reflecting the band structure.}\label{fig.3o}
\end{figure}

We also show in Fig.~\ref{fig.3o}, the Hall conductivity for $N=3$
rhombohedral stacking system calculated based on the Hamiltonian
(\ref{Ham_rhombohedral}). Reflecting the difference of the band
structure (see Fig.~\ref{fig:bands}), the first plateau appears at
$\sigma_{xy}\simeq\pm\omega_{\rm c}\tau\sigma_0$, while
$\sigma_{xy}\simeq\pm 2\omega_{\rm c}\tau\sigma_0$ for Bernal stacking.

\section{Orbital Magnetism and finite-temperature properties}
\label{sec:Orbital_Magnetism}

The orbital magnetism of multilayer graphene with Bernal stacking at
zero temperature has already been discussed by Koshino and
Ando.\cite{Koshino-A_2007b} Here, we turn our attention to the
finite-temperature properties of the magnetic susceptibility.  The
general formula for the orbital magnetic susceptibility of Bloch
electrons was derived by Fukuyama.\cite{Fukuyama_1970} A modified
expression of this formula which is also applicable for the effective
Hamiltonian of Eq.~(\ref{eff_Ham_rhombohedral}) (see
Appendix~\ref{sec:Analytic_results}) is
\begin{equation}
 \chi=-\frac{2}{\beta V}
  \left(\frac{e}{c\hbar}\right)^2
  \sum_{n}\sum_{\bm{k}}
  \tr({\cal G}\gamma_+{\cal G}\gamma_-{\cal G}\gamma_+{\cal G}\gamma_-),
  \label{form_of_susceptibility}
\end{equation}
where $\gamma_{\pm}\equiv(\gamma_x\mp\i\gamma_y)/2$.  This result is
obtained by the Luttinger-Kohn base, the Fourier expansion
$\bm{A}(\bm{r})=\bm{A}_{\bm{q}}
(\e^{\i\bm{q}\cdot\bm{r}}-\e^{-\i\bm{q}\cdot\bm{r}})/2\i$, the $\bm{q}$
expansion of the temperature Green function, and the second derivative
of the thermodynamic potential $\chi=-\frac{1}{V}\left.\frac{\partial^2
\Omega}{\partial B^2}\right|_{B=0}$.  Here, the magnetic field is given
as $B=q_xA_{\bm{q}}^y-q_yA_{\bm{q}}^x=
(\i/2)(q_+A_{\bm{q}}^--q_-A_{\bm{q}}^+)$ with $\bm{q}\to\bm{0}$, where
$A_{\bm{q}}^{\pm}\equiv A_{\bm{q}}^x\pm\i A_{\bm{q}}^y$.

\begin{figure}
 \begin{center}
  \includegraphics[width=78mm]{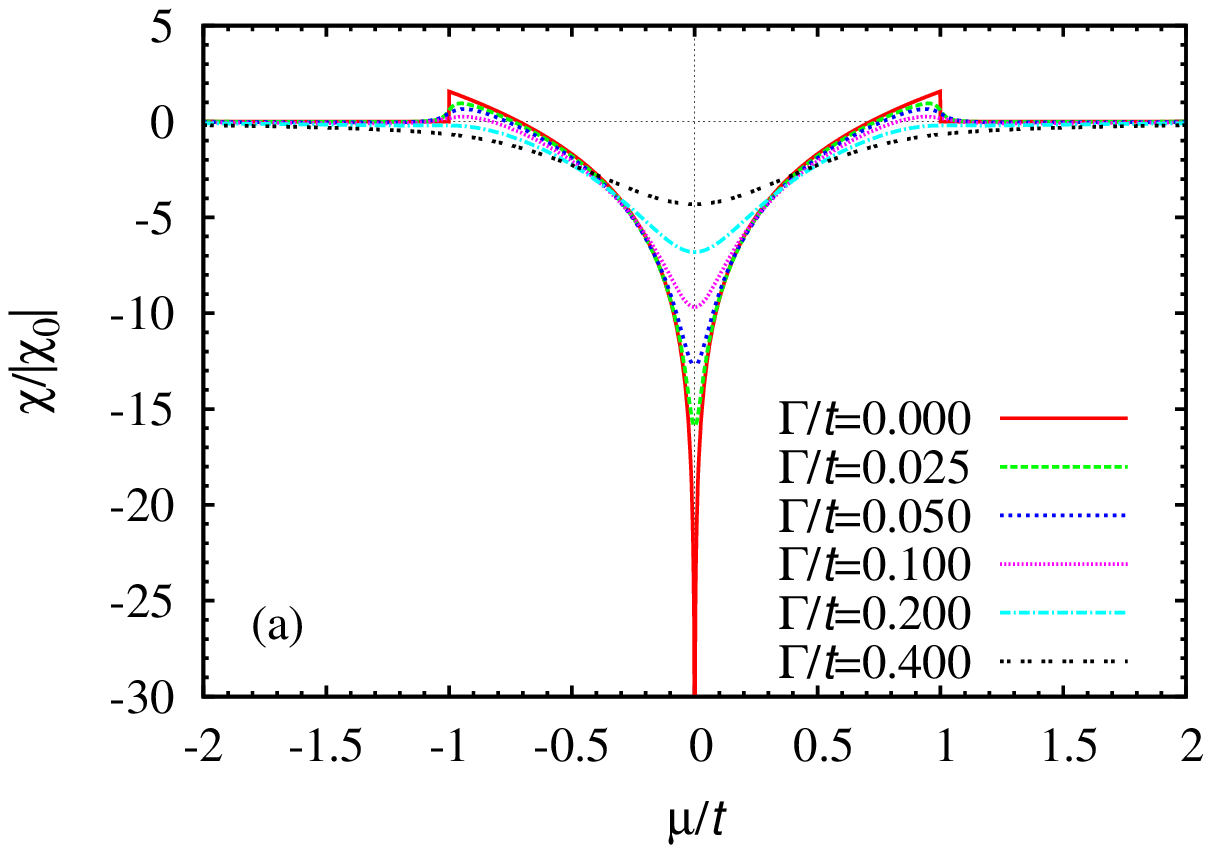}
  \includegraphics[width=75mm]{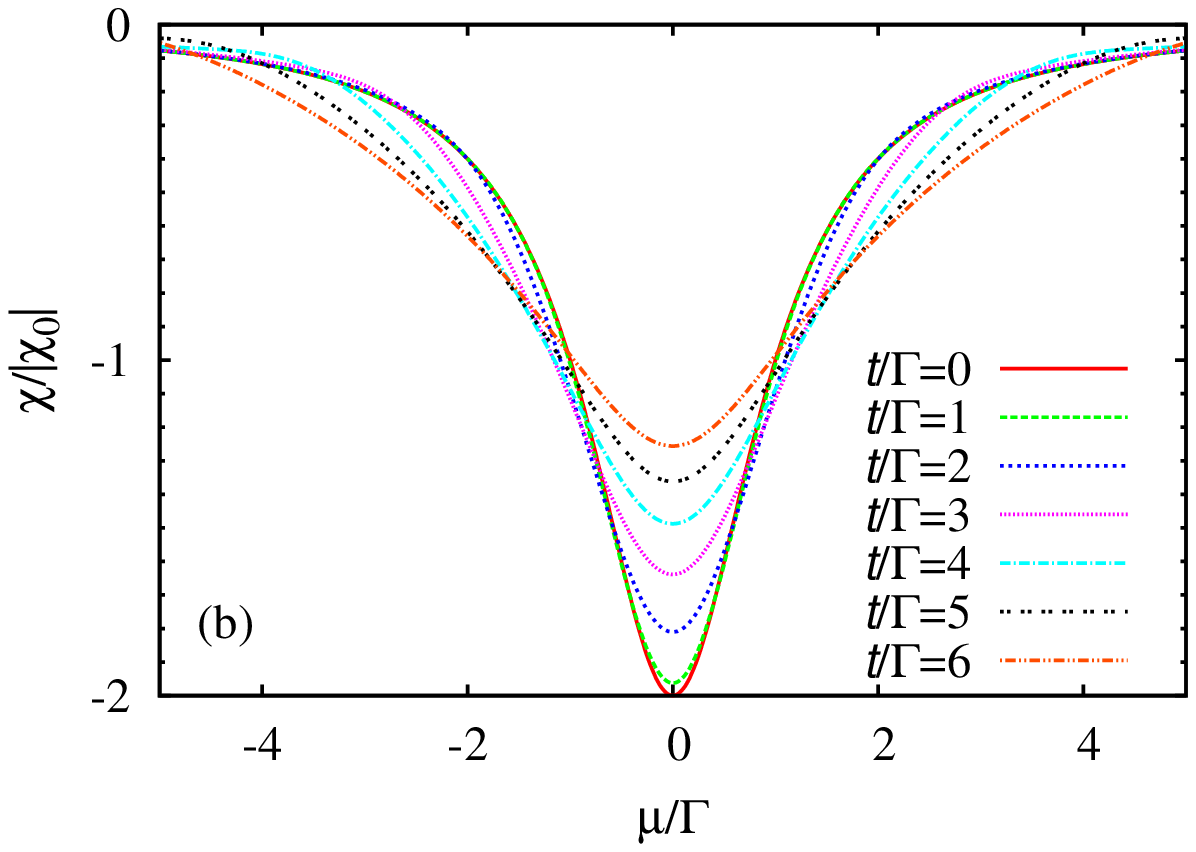}
 \end{center}
 \caption{(Color online) Magnetic susceptibility of the bilayer system
 at zero temperature as a function of the Fermi energy $\mu$ for (a)
 fixed interlayer hopping $t=1$ and for (b) fixed scattering rate
 $\Gamma=1$.  These data are scaled by $\chi_0=-6e^2v^2/\pi^2c^2$. $t$
 dependence of the susceptibility is similar to that of gap dependence
 in the monolayer system [see Fig.~1(a) of
 Ref.~\onlinecite{Nakamura}].}\label{fig.5}
\end{figure}

Figure~\ref{fig.5} shows the susceptibility of the bilayer system at
zero temperature.  (see Appendix~\ref{sec:Analytic_results}) As
discussed by Safran,\cite{Safran} the susceptibility diverges
logarithmically at zero-energy point with $\Gamma=0$, and paramagnetic
regions appear.  As $t/\Gamma$ is increased, the Lorentzian-like curve
becomes broad, and the diamagnetism becomes smaller (larger) at small
(large) Fermi-energy region. This behavior is similar to that of the
monolayer system with an energy gap.\cite{Nakamura} Therefore, as
discussed by Koshino and Ando, odd layer systems with Bernal stacking
tend to show large diamagnetism than the even layer systems due to the
contribution form gapless mode of the effective monolayer.

\begin{figure}
 \begin{center}
  \includegraphics[width=75mm]{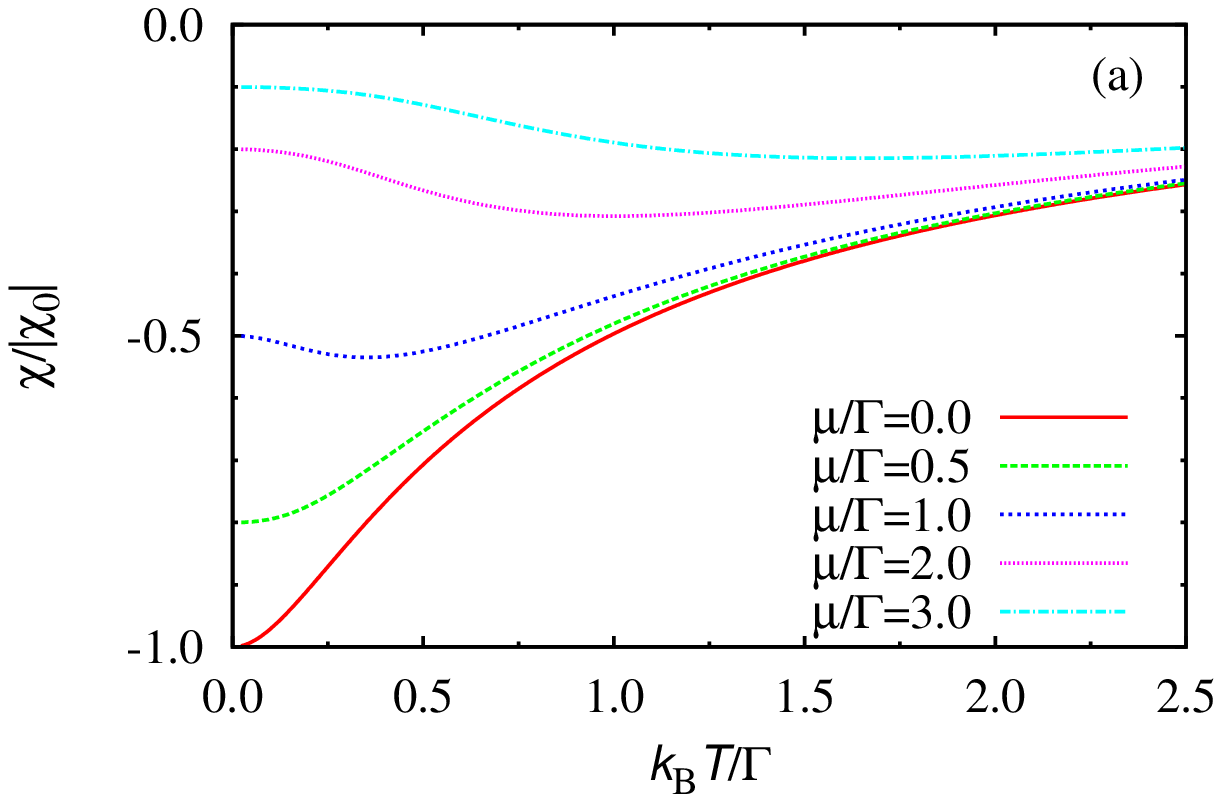}
  \includegraphics[width=75mm]{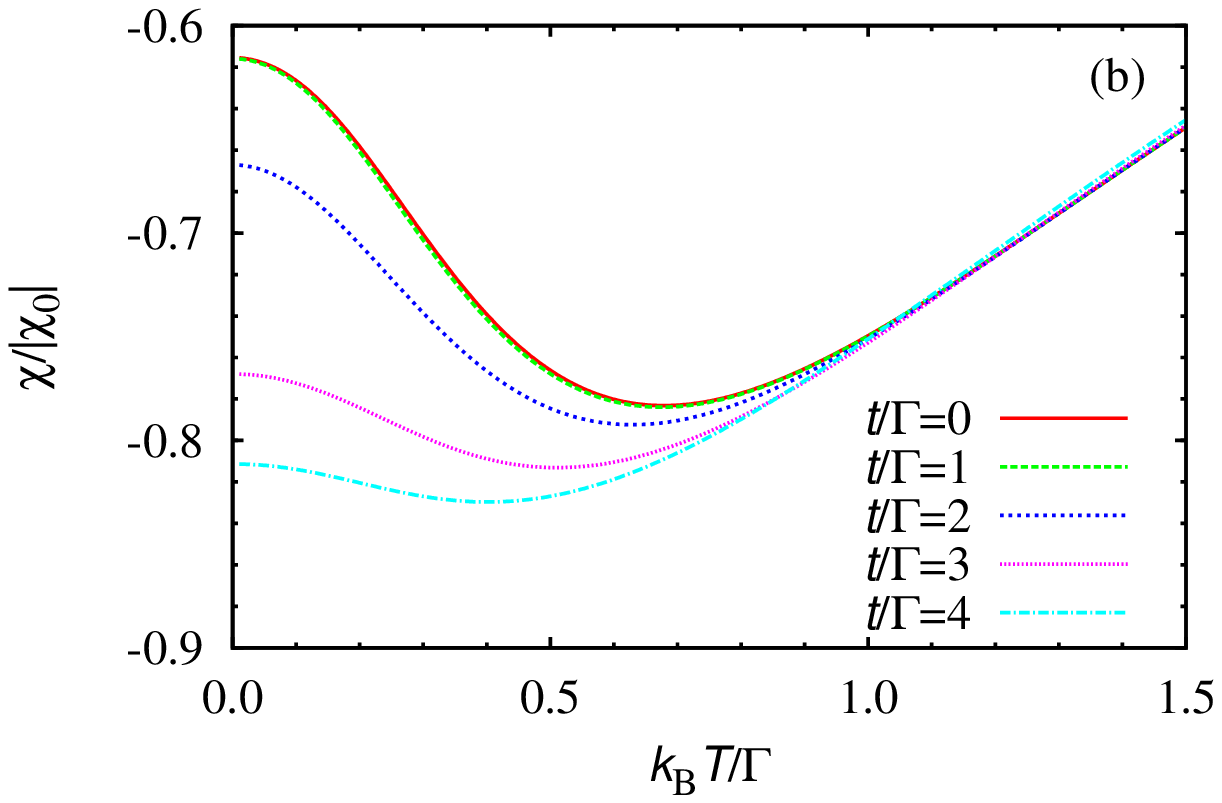}
 \end{center}
 \caption{(Color online) Temperature dependence of the magnetic
 susceptibility $\chi$ of (a) the monolayer system and (b) the bilayer
 system for several strength of the interlayer hopping at
 $\mu/\Gamma=1.5$. A minimum value appears as a function of
 temperature.}\label{fig.6}
\end{figure}

\begin{figure}
 \begin{center}
  \includegraphics[width=75mm]{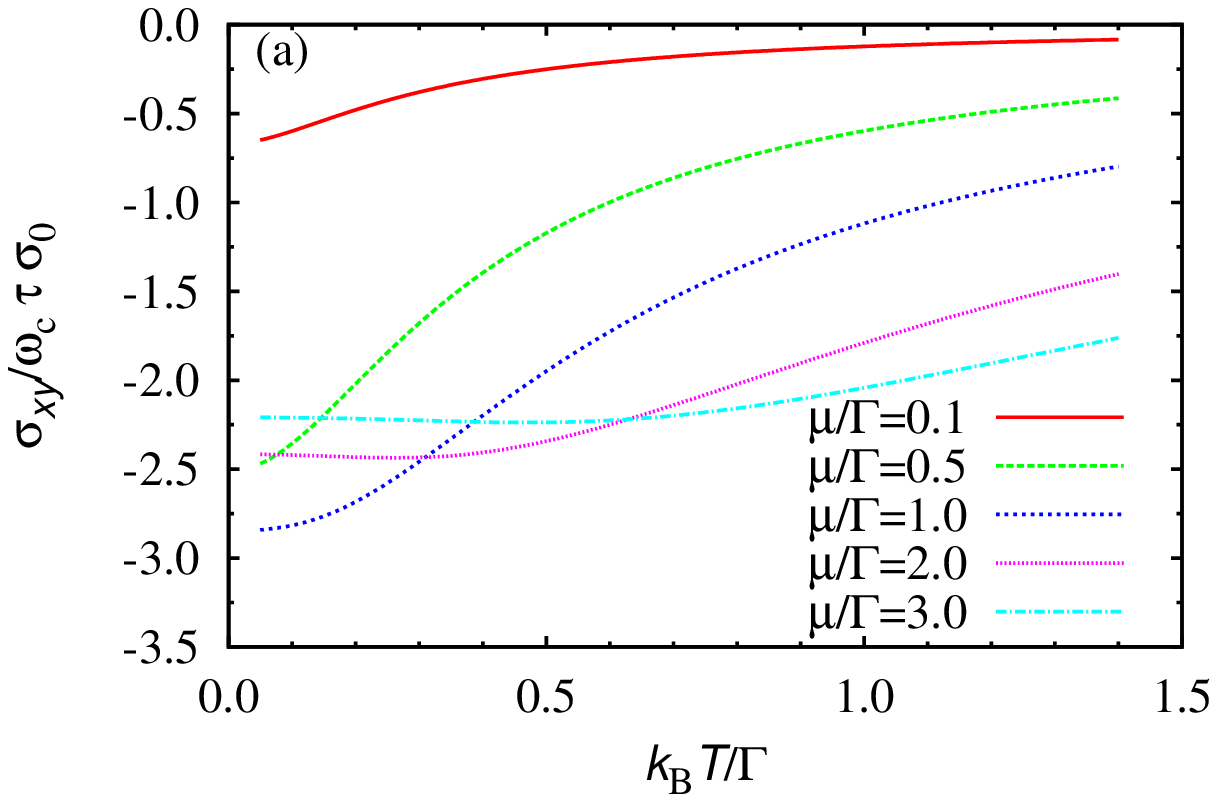}
  \includegraphics[width=75mm]{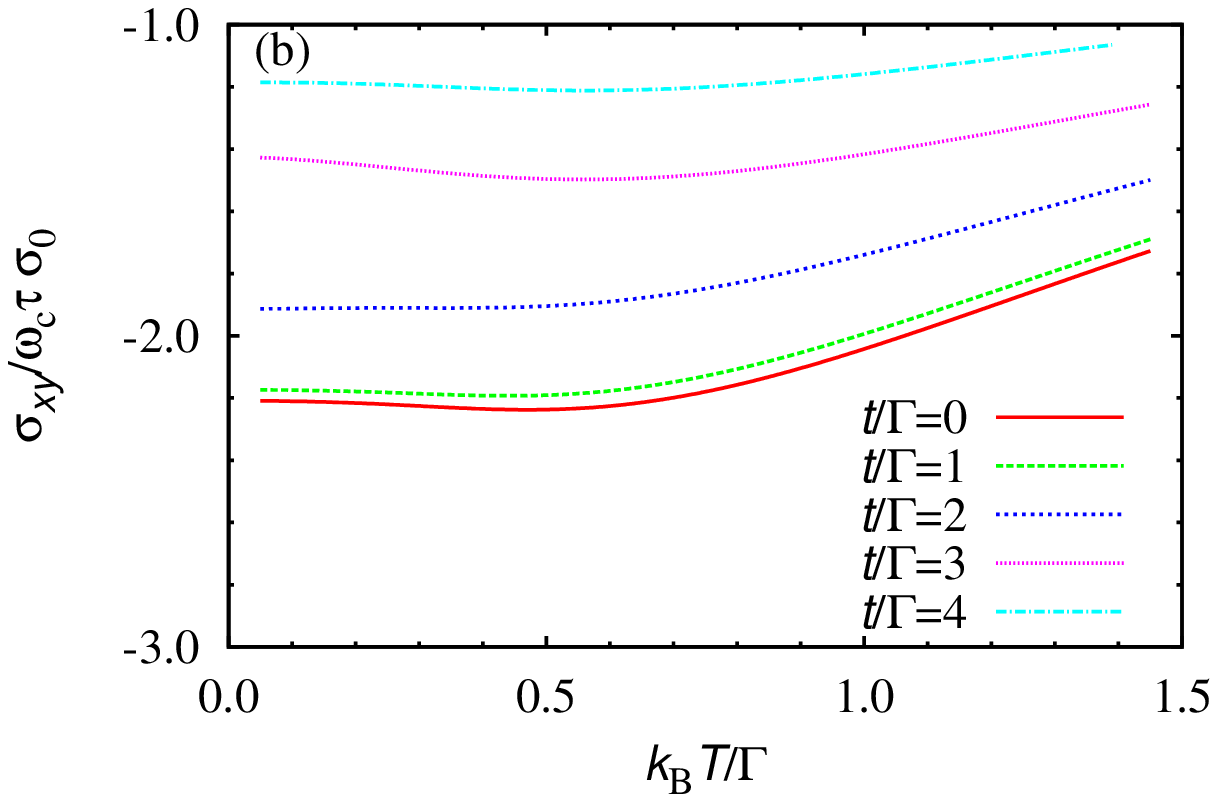}
 \end{center}
 \caption{(Color online) Temperature dependence of the Hall conductivity
 $\sigma_{xy}$ of (a) the monolayer system and (b) the bilayer system
 for several strength of the interlayer hopping at $\mu/\Gamma=3$.  A
 minimum value appears as a function of temperature.}\label{fig.7}
\end{figure}

\begin{figure}
 \begin{center}
  \includegraphics[width=75mm]{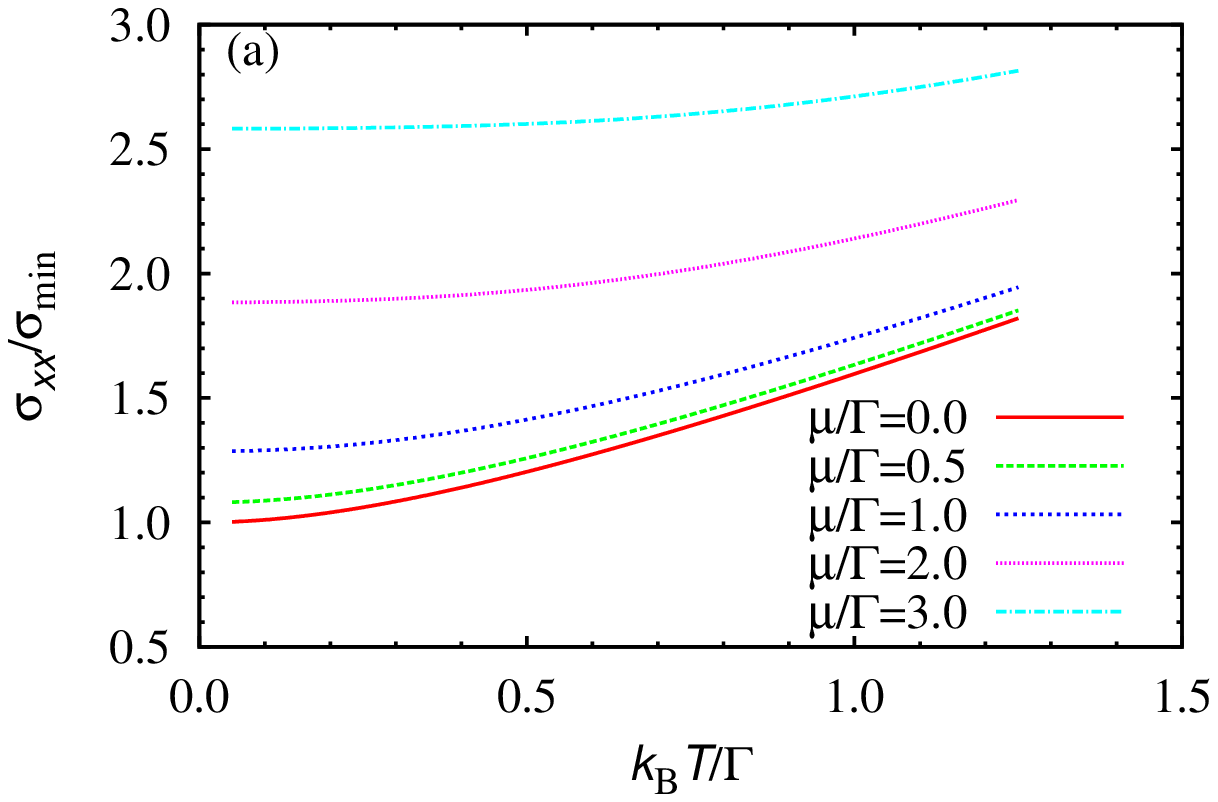}
  \includegraphics[width=75mm]{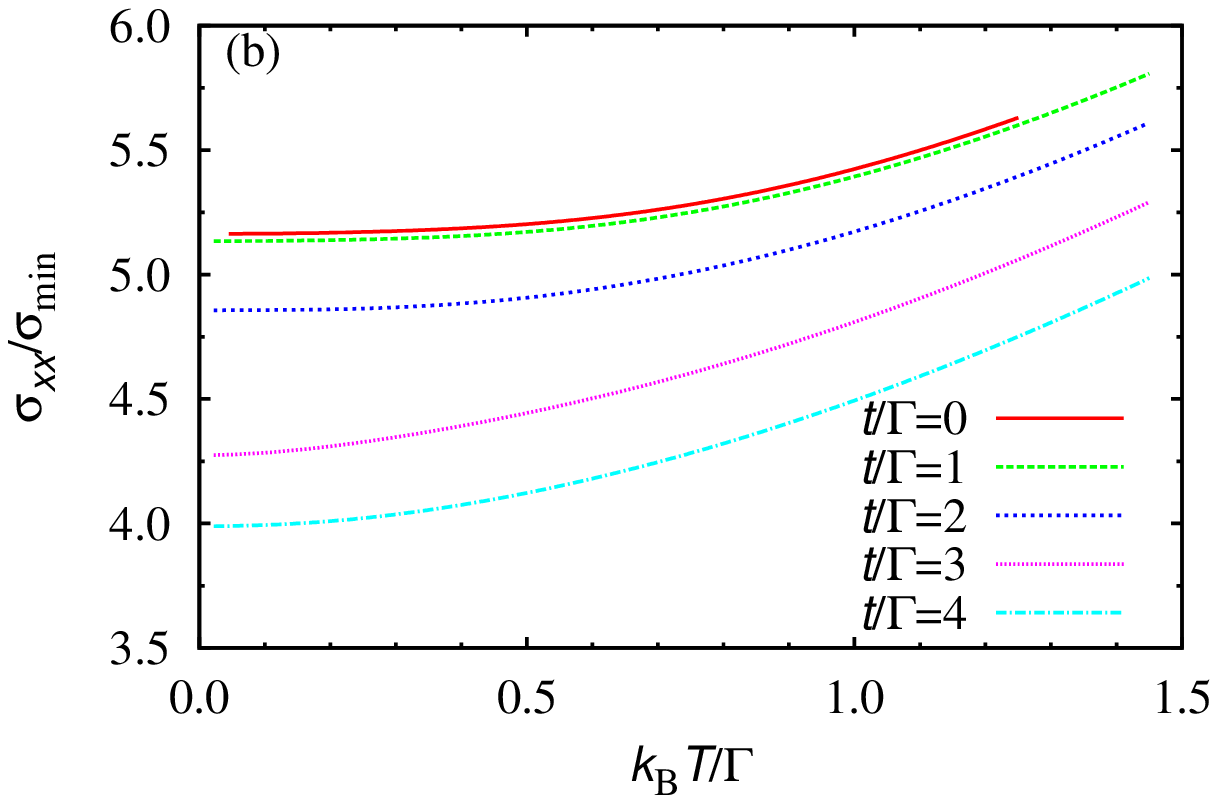}
 \end{center}
 \caption{(Color online) Temperature dependence of the conductivity
 $\sigma_{xx}$ of (a) the monolayer system and (b) the bilayer system
 for several strength of the interlayer hopping at $\mu/\Gamma=3$.  They
 behave monotonically as a function of temperature.}\label{fig.8}
\end{figure}

In Fig.~\ref{fig.6}, we show the temperature dependence of the
susceptibility of monolayer and bilayer systems. We find that there
appears the minimum value at finite temperature when the gate voltage
satisfies the relation with some critical value $|\mu|>\mu_{\rm
c}$. This phenomenon is explained in the following way: For simplicity,
let us consider the susceptibility in the monolayer system with finite
$\Gamma$,\cite{Fukuyama_2007}
\begin{equation}
 \chi=-\frac{e^2 v^2}{6 \pi^2 c^2}
  \int_{-\infty}^{\infty}\d x f(x)\Im\frac{1}{(x+\i\Gamma)^2},
\label{chi_mono}
\end{equation}
where $f(x)\equiv(\e^{\beta(x-\mu)}+1)^{-1}$ is the Fermi distribution
function.  At zero and high temperature limits, Eq.~(\ref{chi_mono})
becomes $\chi=-\frac{e^2 v^2}{6\pi^2 c^2}\frac{\Gamma}{\mu^2+\Gamma^2}$
and $\chi=0$, respectively.  In the finite-temperature regions,
Sommerfeld expansion of (\ref{chi_mono}) is
\begin{equation}
 \frac{6\pi^2 c^2}{e^2 v^2}\chi
  =-\frac{\Gamma}{\mu^2+\Gamma^2}
  +\frac{\pi^2}{6}(k_{\rm B} T)^2 F'(\mu)
  +O(k_{\rm B} T)^3,
\end{equation}
where
\begin{equation}
 F(x)\equiv\Im\frac{1}{(x+\i\Gamma)^2},\quad
 F'(x)=\frac{-2\Gamma\left[3x^2-\Gamma^2\right]}{\left[x^2+\Gamma^2\right]^3}.
\end{equation}
Therefore, for $F'(\mu)<0$ ($|\mu|/\Gamma>1/\sqrt{3}$) a minimum value
appears as a function of temperature.  This means that the condition to
appear a minimum value is existence of an inflexion point in the
susceptibility at zero temperature as a function of the Fermi-energy
$\propto \int_{-\infty}^{\mu}F(x)\d x$.  The same argument can also be
applied to multilayer systems and other physical quantities, and it
turns out that the Hall conductivity may have minimum value for small
interlayer hopping or odd layer cases (see Fig.~\ref{fig.7}), where the
peak structure of the Hall conductivity is clear.  Similar argument for
a minimum value of physical quantity at finite temperature is recently
done for the magnetization of quantum spin chains in a magnetic
field.\cite{Maeda-H-O} For the longitudinal conductivity, there appears
no minimum (see Fig.~\ref{fig.8}).

\section{Summary and discussion}
In summary, we have studied the electric transport and the orbital
magnetism of multilayer graphene in weak-magnetic field. We have found
that kinks appear in the gate voltage dependence of the conductivity,
and plateaux in the Hall conductivity. These phenomena are explained as
multiband effects which become clear when the energy gap and the
collision time satisfy the condition $\Delta\gg \hbar/\tau$.  We have
also considered finite-temperature properties of this system, and found
that a minimum value appears in the magnetic susceptibility and the Hall
conductivity, as functions of temperature. These phenomena are explained
by the existence of an inflexion point in the zero-temperature
Fermi-energy dependence.

In this paper, we have tuned our attention mainly to the Bernal stacking
systems, except for the result of Fig.~\ref{fig.3o}.  For rhombohedral
staking systems, the band structure is quite different from that of the
Bernal stacking, especially the gapless linear dispersions for odd
$N\geq 3$ do not appear.  Therefore, it is expected that difference of
the properties of the physical quantities, such as the magnetic
susceptibility, is not so drastic depending on the parity of the number
of layers. It is also reported that the trigonal wrapping effect causes
essential difference in the longitudinal conductivity.\cite{Cserti-C-D}
This would be also an interesting future problem. We have also discussed
the semiclassical analysis of transport properties using the results of
Boltzmann equation for nonrelativistic electrons.  It is expected,
however, that Boltzmann equation for Dirac type systems is needed for
more detailed description in regions close to the zero-energy point.
Moreover, it is also desirable to extend the present analysis to
finite-magnetic field regions where the Landau quantization is
essential.

Note added in proof: After the submission of this paper, we become aware
of a preprint H. Min and A.H. MacDonald, arXiv:0711.4333 where the
present matrix decomposition technique and the 2 $\times$ 2 effective
Hamiltonian are used to discuss the quantum Hall effect in multilayer
graphenes with general stacking structures.

\section{ACKNOWLEDGMENTS}

We are grateful to M.~Oshikawa and A.~Tokuno for discussions.
M.~N. thanks T.~Ando and A.~H.~MacDonald for valuable comments.

\appendix

\section{Derivation of the effective Hamiltonian}\label{sec:eff_Ham}

In order to calculate physical quantities of multilayer systems, it is
useful to reduce the original Hamiltonian with $2N\times 2N$ matrix form
into some effective Hamiltonian which has fewer matrix elements.  McCann
and Fal'ko reduced the $4\times 4$ matrix Hamiltonian into the $2\times
2$ form to discuss the quantum Hall effect of the bilayer system, which
describes the two bands near the zero-energy point.\cite{McCann-F} In
order to derive such effective Hamiltonian, we should find ${\cal
H}_{\rm eff}$ which have the same eigenvalues of the original
Hamiltonian ${\cal H}$:
\begin{equation}
 \det(\varepsilon-{\cal H})=\det(\varepsilon-{\cal H}_{\rm eff})=0.
  \label{determinants}
\end{equation}

For the rhombohedral $N$-layer system (\ref{Ham_rhombohedral}), the
$2\times 2$ effective Hamiltonian is obtained in the following way:
First, we change the order of the matrix elements from $A_1, B_1, A_2,
B_2, \dots$ to $\dots B_3, A_2, B_1, A_1, B_2, A_3, \dots$, where $A_i$
and $B_i$ indicate two sublattices of the hexagonal lattice of $i$th
layer. Then, we have
\begin{equation}
 {\cal H}=
\left[
  \begin{array}{c cc cc cc cc c}
 \ddots &t   &    &    &    &    &    &    &    &   \\
   t   &    &    &    &    &    &    &    &k_- &    \\
       &    &    &t   &    &    &    &k_+ &    &    \\
       &    &t   &    &    &    &k_- &    &    &    \\
       &    &    &    &    &k_+ &    &    &    &    \\
       &    &    &    &k_- &    &t   &    &    &    \\
       &    &    &k_+ &    &t   &    &    &    &    \\
       &    &k_- &    &    &    &    &    &t   &    \\
       &k_+ &    &    &    &    &    &t   &    &    \\
       &    &    &    &    &    &    &    &    &\ddots \\
  \end{array}
\right],
\end{equation}
where $k_{\pm}\equiv k_x\pm\i k_y$ is eigenvalue of the momentum
operator $\pi_+$.  We have set $v=\hbar=1$ for simplicity. Next, we
calculate $\det(\varepsilon-{\cal H})$ approximately for $\varepsilon\ll
t$, then we obtain
\begin{equation}
 \det(\varepsilon-{\cal H})\simeq
  \varepsilon^2t^{2N-2}-k^{2N}=0.
  \label{determinants.2}
\end{equation}
One of effective Hamiltonians which satisfies Eqs.~(\ref{determinants})
and (\ref{determinants.2}) can be chosen as
Eq.~(\ref{eff_Ham_rhombohedral}).

On the other hand, the effective Hamiltonian for the Bernal stacking
graphenes is discussed by Koshino and Ando, quite
recently.\cite{Koshino-A_2007b} According to their result, $N$-layer
Bernal stacking system can be described by isolated $[N/2]$ bilayer
systems with some effective interlayer hopping, and one monolayer system
if $N$ is odd. This is exact mapping of Eq.~(\ref{Ham_Bernal}) without
using any approximation. Here, we derive the same result by the argument
based on the determinants. After the same reordering of the original
Hamiltonian (\ref{Ham_Bernal}) as the rhombohedral stacking, we have
\begin{equation}
 {\cal H}=
 \left[
   \begin{array}{cccccccc}
    \ddots&&&&&&&\cdot\\
         &    &    &    &    &    & k_+&   \\
         &    &    &    &    & k_-&    &   \\
         &    &    &    & k_+&    &    &   \\
         &    &    & k_-&    &  t &    &   \\
         &    & k_+&    &  t &    &  t &   \\
         & k_-&    &    &    &  t &    & t \\
    \cdot&    &    &    &    &    &  t &\ddots
   \end{array}
 \right].
 \label{Ham_Bernal.2}
\end{equation}
Then we obtain the following recursion relation for the determinant of
the $N$-layer system $A_N\equiv \det(\varepsilon-{\cal H})$ as
\begin{align}
 A_1&=\varepsilon^2-k^2,\\
 A_2&=(\varepsilon^2-k^2)^2- t^2\varepsilon^2,\\
 &\cdots\nonumber\\
 A_N&
 =(\varepsilon^2-k^2)
 A_{N-1}-t^2\varepsilon^2 A_{N-2}.
\label{recursion.2}
\end{align}
Note that recursion relation such a closed form cannot be obtained for
the rhombohedral stacking.  Equation~(\ref{recursion.2}) can be solved
exactly, and factorized in the following way:
\begin{align*}
 A_1&\equiv X,\\
 A_2&=X^2-t^2\varepsilon^2,\\
 A_3&=X(X^2-2t^2\varepsilon^2),
 \\
 A_4&=
 \left[X^2-\left(\textstyle\frac{\sqrt{5}-1}{2}\right)^2t^2
 \varepsilon^2\right]
 \left[X^2-\left(\textstyle\frac{\sqrt{5}+1}{2}\right)^2t^2
 \varepsilon^2\right],\\
 A_5&=X(X^2-t^2\varepsilon^2)(X^2-3t^2\varepsilon^2),
\\
&\cdots.
\end{align*}
Thus, the determinant for the $N$-layer system can be decomposed into
those of the bilayers $[X^2-(t^*)^2\varepsilon^2]$ with the effective
hopping $t^*$ (Table~\ref{EILH}) and that of one monolayer $X$ for odd
$N$. This means that the original Hamiltonian (\ref{Ham_Bernal}) can be
block diagonalized into subsystems.
Koshino and Ando have obtained the effective hopping as
$t^*=2t\sin[m\pi/2(N+1)]$ with $m$ being an appropriate integer, by
considering diagonalization of the matrix
(\ref{Ham_Bernal.2}).\cite{Koshino-A_2007b} This factorization of the
Hamiltonian is analogous to the $N$-leg ladder in quantum spin systems
where a flat dispersion appears for odd $N$.\cite{Sato-O}

\section{Analytic results}\label{sec:Analytic_results}
Here, we present anaclitic forms of physical quantities discussed in
this paper. We also briefly discuss the results based on the low-energy
effective Hamiltonian of the rhombohedral stacking
(\ref{eff_Ham_rhombohedral}) given by $2\times 2$ matrix.

The longitudinal conductivity is given by the following form,
\begin{equation}
 \sigma=\frac{e^2}{4 \pi^2\hbar}
  \int_{-\infty}^{\infty}\d x
  [-f'(x)]{\cal A}_N(x),
  \label{conductivity.5}
\end{equation}
where $f(x)=(\e^{\beta(x-\mu)}+1)^{-1}$ .  For the monolayer case $N=1$,
we have
\begin{equation}
 {\cal A}_1(x)=
  \frac{\Gamma^2+x^2}{\Gamma x}\arctan\left(\frac{x}{\Gamma}\right)+1.
  \label{conductivity.6}
\end{equation}
For the bilayer system $N=2$ with $4\times 4$ Hamiltonian, the result is
\begin{widetext}
\begin{align}
&\lefteqn{
 {\cal A}_2(x)=
  -\frac{t x^2}{\Gamma\left(t^2-4 x^2\right)}
 \left\{\frac{\pi}{2} 
 - \arctan\left(\frac{-t^2+\Gamma ^2+x^2}{2 t \Gamma}\right)\right\}
 +\frac{ t \Gamma ^2}{(t^2+4 \Gamma ^2)x}
 \arctanh\left(\frac{2tx}{t^2+\Gamma^2+x^2}\right)}\nonumber\\
&
 +\frac{x^2+\Gamma ^2}{2\Gamma  x}
 \left(\frac{t^4}{\left(t^2-4 x^2\right)
 \left(t^2+4 \Gamma ^2\right)}+1\right) 
 \left\{2\arctan\left(\frac{x}{\Gamma }\right)
 -\arctan\left(\frac{t^2+\Gamma ^2-x^2}{2 x \Gamma }\right)
 +\frac{\pi}{2} \sgn(x)\right\}+2.
\end{align}
\end{widetext}

For the $2\times 2$ Hamiltonian (\ref{eff_Ham_rhombohedral}),
$\sigma_{xx}$ is given by $N$ times of Eq.~(\ref{conductivity.5}) with
Eq.~(\ref{conductivity.6}). This is consistent with the result of
Ref.~\onlinecite{Cserti} where $N=2$ and $\Gamma=\mu=0$ case is
discussed.

We omit analytical result of the Hall conductivity $\sigma_{xy}$ for the
bilayer system, because of its lengthiness.  For $\sigma_{xy}$ of the
$2\times 2$ Hamiltonian (\ref{eff_Ham_rhombohedral}), the following two
terms should be added in $\tr(\cdots)$ of the polarization function
(\ref{Pi_CC4}), since $\gamma_{\mu}$ has momentum dependence,
\begin{equation}
 -{\cal G}\gamma_x{\cal G}_+\gamma_y{\cal G}_+(\partial_{k_x}\gamma_y)
  +{\cal G}\gamma_x{\cal G}_+(\partial_{k_x}\gamma_y){\cal G}\gamma_y.
  \label{im-time_Pi.9x}
\end{equation}
In this case, finite cutoff is needed to make the $\sigma_{xy}$
finite. Moreover, the Drude-Zener-like behavior for the high-energy
region cannot be described. As a result, it turns out that Hamiltonian
(\ref{eff_Ham_rhombohedral}) is not appropriate to discuss the Hall
conductivity in a weak-magnetic field.


The magnetic susceptibility (\ref{form_of_susceptibility}) for bilayer
system with $4\times 4$ Hamiltonian is obtained as
\begin{equation}
 \chi=
  \frac{v^2 e^2}{12\pi^2 c^2}
  \int_{-\infty}^{\infty}\d x f(x){\cal B}_{2}(x),
  \label{sus_bilayer.1}
\end{equation}
and
\begin{align}
{\cal B}_{2}(x)=&
 \Im
 \biggl[
  \frac{2}{(x+\i\Gamma)^2-t^2}\nonumber\\
&
 -\frac{3}{t(x+\i\Gamma)}\log\left[\frac{t+x+\i\Gamma}{-t+x+\i\Gamma}\right]
\biggr].
\end{align}
In the limit $\Gamma\to 0$, Eq.~(\ref{sus_bilayer.1}) becomes the result
obtained by Safran,\cite{Safran}
\begin{align}
\chi
 =&\frac{v^2e^2}{12\pi c^2 t}
 \left[f(-t)-f(t)
 +3{\cal P}\int_{-t}^{t}\d x\frac{f(x)}{x} \right]\\
 =&
 \frac{v^2e^2}{12\pi c^2 t}\theta(t-|\mu|)
 \left(1+3\log\left|\frac{\mu}{t}\right|\right)\quad (T=0),
\end{align}
where ${\cal P}$ means the Cauchy's principal value.  On the other hand,
in the limit $t\to 0$, we obtain a value two times of that of the
monolayer system (\ref{chi_mono}).

The derivation of the formula (\ref{form_of_susceptibility}) for the
$2\times 2$ Hamiltonian (\ref{eff_Ham_rhombohedral}) can be done as in
the same way of Ref.~\onlinecite{Fukuyama_1970}. In this calculation, we
should extract terms of the thermodynamic potential which is
proportional to squire of the magnetic field in the following
representation:
\begin{equation}
 B^2=\frac{2q_+q_-A_+A_--q_+^2A_-^2-q_-^2A_+^2}{4}.
\end{equation}
The formula for $N=2$ was also derived in
Ref.~\onlinecite{Koshino-A_2007b} using $\gamma_x$, $\gamma_y$ which has
more complicated expression. These two expressions can be shown to be
equivalent using some identities. To calculate susceptibility for
Eq.~(\ref{eff_Ham_rhombohedral}), a finite cutoff is needed to make the
susceptibility finite. This result shows diamagnetism near the zero
energy point, but it is not appropriate for quantitative analysis.




\begin{thebibliography}{99}

 \bibitem{Novoselov}
 K.~S.~Novoselov, A.~K.~Geim, S.~V.~Morozov,
 D.~Jiang, M.~I.~Katsnelson, I.~V.~Grigorieva, S.~V.~Dubonos, and
 A.~A.~Firsov, Nature (London) {\bf 438}, 197 (2005).
 
 \bibitem{Zhang}
 Y.~Zhang, Y.-W.~Tan, H.~L.~Stormer, and P.~Kim,
 Nature (London) {\bf 438}, 201 (2005).

 \bibitem{Ludwig-F-S-G}
 A.~W.~Ludwig, M.~P.~A.~Fisher, R.~Shankar, and G.~Grinstein,
 Phys. Rev. B {\bf 50}, 7526 (1994).

 \bibitem{Shon-A}
 N.~H.~Shon and T.~Ando,
 J.~Phys.~Soc.~Jpn. {\bf 67}, 2421 (1998).

 \bibitem{Zheng-A}
 Y.~Zheng and T.~Ando, Phys. Rev. B {\bf 65}, 245420 (2002).

 \bibitem{Gusynin-S_2005b}
 V.~P.~Gusynin and S.~G.~Sharapov,
 Phys. Rev. Lett. {\bf 95}, 146801 (2005).
 
 \bibitem{Gusynin-S_2006}
 V.~P.~Gusynin and S.~G.~Sharapov,
 Phys. Rev. B {\bf 73}, 245411 (2006).

 \bibitem{Peres-G-C}
 N.~M.~R.~Peres, F.~Guinea, and A.~H.~Castro Neto,
 Phys. Rev. B {\bf 73}, 125411 (2006).

 \bibitem{Ziegler_2007}
 K.~Ziegler, Phys. Rev. B {\bf 75}, 233407 (2007), and references therein.

 \bibitem{McCann-F}
 E. McCann and V. I. Fal'ko,
 Phys. Rev. Lett. {\bf 96}, 086805 (2006).

 \bibitem{Nilsson-C-G-P}
 J.~Nilsson, A.~H.~Castro Neto, F.~Guinea, and N.~M.~R.~Peres,
 Phys. Rev. Lett. {\bf 97}, 266801 (2006).

 \bibitem{Koshino-A_2006}
 M. Koshino and T. Ando,
 Phys. Rev. B {\bf 73}, 245403 (2006).

 \bibitem{Cserti}
 J. Cserti, Phys. Rev. B {\bf 75}, 033405 (2007).

 \bibitem{Cserti-C-D}
 J.~Cserti, A.~Csordas, and G.~David,
 Phys. Rev. Lett. {\bf 99}, 066802 (2007).

 \bibitem{Ohta}
%
%
	 T.~Ohta, A.~Bostwick, J.~L.~McChesney, T.~Seyller,
	 K.~Horn, and E.~Rotenberg,
	 Phys. Rev. Lett. {\bf 98}, 206802 (2007).

 \bibitem{Bostwick}
 A.~Bostwick, T.~Ohta, J.~L.~McChesney, K.~V.~Emtsev, T.~Seyller,
 K.~Horn, and E.~Rotenberg, arXiv:0705.3705 

 \bibitem{Zhou}
 S.~Y.~Zhou, G.-H.~Gweon, A.~V.~Fedorov, P.~N.~First, W.~A.~de Heer,
 D.-H.~Lee, F.~Guinea, A.~H.~Castro Neto, and  A.~Lanzara,
 Nature Materials {\bf 6}, 770 (2007).

 \bibitem{Partoens-P}
 B.~Partoens and F.~M.~Peeters,
 Phys. Rev. B {\bf 74}, 075404 (2006)


 \bibitem{Latil-H}
 S.~Latil and L.~Henrard, Phys. Rev. Lett. {\bf 97}, 036803 (2006).

 \bibitem{Guinea-C-P}
 F.~Guinea, A.~H.~Castro Neto, and N.~M.~R. Peres,
 Phys. Rev. B {\bf 73}, 245426 (2006).

 \bibitem{Manes-G-V}
 J.~L.~Ma\~{n}es, F.~Guinea, and Maria A.~H.~Vozmediano,
 Phys. Rev. B {\bf 75}, 155424 (2007)

 \bibitem{McClure}
 J.~W.~McClure, Phys. Rev. {\bf 104}, 666 (1956).

 \bibitem{Safran-D}
 S.~A.~Safran and F.~J.~DiSalvo,
 Phys. Rev. B {\bf 20}, 4889 (1979).

 \bibitem{Ghosal-G-C}
 A.~Ghosal, P.~Goswami, and S.~Chakravarty,
 Phys. Rev. B {\bf 75}, 115123 (2007).

 \bibitem{Fukuyama_2007}
 H. Fukuyama, J. Phys. Soc. Jpn {\bf 76}, 043711 (2007).

 \bibitem{Koshino-A_2007a}
 M. Koshino and T. Ando,
 Phys. Rev. B {\bf 75}, 235333 (2007).

 \bibitem{Nakamura}
 M. Nakamura, Phys. Rev. B {\bf 76}, 113301 (2007).

 \bibitem{Safran}
 S.~A.~Safran, Phys. Rev. B {\bf 30}, 421 (1984).

 \bibitem{Saito-K}
 R.~Saito and H.~Kamimura, Phys. Rev. B {\bf 33}, 7218 (1986).

 \bibitem{Koshino-A_2007b}
 M. Koshino and T. Ando,
 Phys. Rev. B {\bf 76}, 085425 (2007).

 \bibitem{Mahan}
	 For example, see G. D. Mahan, {\it Many-Particle Physics}
	 (Plenum, New York, 2000).

 \bibitem{Abrikosov-G-D}
 For example see,
 A. A. Abrikosov, L. P. Gorkov, and I. E. Dzyaloshinski, 1965,
 {\it Methods of Quantum Field Theory in Statistical Physics},
 (Dover, New York).

 \bibitem{Nomura-M}
 K.~Nomura and A.~H.~MacDonald,
 Phys. Rev. Lett. {\bf 98}, 076602 (2007).

 \bibitem{Peres-LS-S}
 N.~M.~R.~Peres, J.~M.~B.~Lopes dos Santos, and T.~Stauber,
 Phys. Rev. B {\bf 76}, 073412 (2007).

 \bibitem{Luttinger-K}
 J.~M.~Luttinger and W.~Kohn,
 Phys. Rev. {\bf 97}, 869 (1955).

 \bibitem{Fukuyama-E-W}
 H. Fukuyama, H. Ebisawa, and Y. Wada,
 Prog. Theor. Phys. {\bf 42}, 494 (1969).

 \bibitem{Fukuyama_2006}
 H. Fukuyama, Ann. Phys. (Leipzig) {\bf 15}, 520 (2006).

 \bibitem{Yang-N}
 X. Yang and C. Nayak,
 Phys. Rev. B {\bf 65}, 064523 (2002).

 \bibitem{Morozov}
 S. V. Morozov, K. S. Novoselov, F. Schedin, D. Jiang,
 A. A. Firsov, and A. K. Geim, Phys. Rev. B {\bf 72}, 201401(R) (2005).

 \bibitem{Fukuyama_1970}
 H. Fukuyama, Prog. Theor. Phys. {\bf 45}, 704 (1971).

 \bibitem{Maeda-H-O}
 Y.~Maeda, C.~Hotta, and M.~Oshikawa,
 Phys. Rev. Lett. {\bf 99}, 057205 (2007).

 \bibitem{Sato-O}
 For example, see
 M.~Sato and M.~Oshikawa,
 Phys. Rev. B {\bf 75}, 014404 (2007).
\end{thebibliography}
\end{document}